\documentclass{article}
\usepackage[margin=2cm, includefoot, footskip=30pt]{geometry}

\usepackage{amsmath}
\usepackage{amsthm}
\usepackage{amssymb}
\usepackage{hyperref}
\usepackage{standalone}
\usepackage{subcaption}
\usepackage{adjustbox}
\usepackage{tikz}
\usepackage{booktabs}
\usepackage{multicol,multirow,array}
\usepackage{graphicx}
\usepackage{algorithm,algorithmic}
\usepackage{authblk}
\usepackage{tabularx}
\usepackage{color, colortbl}
\usepackage{smartdiagram}
\usepackage[round]{natbib}
\usesmartdiagramlibrary{additions}
\usetikzlibrary{er,positioning, calc, patterns}
\usetikzlibrary{decorations.pathreplacing}

\definecolor{Gray}{gray}{0.92}
\definecolor{background}{RGB}{5, 66, 81}

\theoremstyle{definition}

\newcommand{\totalarticles}{2422}
\newcommand{\manual}{76}
\newcommand{\authors}{4226}
\newcommand{\edges}{7642}

\newcommand{\isolatedpercentage}{8.0}
\newcommand{\connectedcomponents}{1157}

\newcommand{\largestcc}{796}

\title{A bibliometric study of research topics, collaboration and centrality 
       in the Iterated Prisoner's Dilemma}

\author[1, 2, *]{Nikoleta E. Glynatsi}
\author[1]{Vincent A. Knight}

\affil[1]{Cardiff University, School of Mathematics, United Kingdom}
\affil[2]{Max Planck Institute for Evolutionary Biology, Germany}
\affil[*]{Corresponding author: Nikoleta E. Glynatsi, glynatsi@evolbio.mpg.de}
\date{}
\begin{document}

\maketitle

\begin{abstract}
This manuscript explores the research topics and collaborative behaviour of
authors in the field of the Prisoner's Dilemma using topic modeling and a graph
theoretic analysis of the co-authorship network. The analysis identified five
research topics in the Prisoner's Dilemma which have been relevant over the course
of time. These are human subject research, biological studies, strategies,
evolutionary dynamics on networks and modeling problems as a Prisoner's Dilemma
game. Moreover, the results demonstrated the Prisoner's Dilemma is a field of
continued interest, and that it is a collaborative field compared to other
game theoretic fields. The co-authorship
network suggests that authors are focused on their communities and that not many
connections across the communities are made. The most central authors of the network
are the authors connected to the main cluster. Through examining the networks
of topics, it was uncovered that the main cluster is characterised by the
collaboration of authors in a single topic.

These findings add to the bibliometrics study in another field and present new questions 
and avenues of research to understand the reasons for the measured behaviours.
\end{abstract}

\section{Introduction}\label{section:introduction}

The Prisoner's Dilemma (PD) is a well known game used since its introduction in
the 1950's~\citep{Flood1958} as a framework for studying the emergence of
cooperation; a topic of continued interest for mathematical,
social, biological and
ecological sciences. This manuscript presents a bibliometric
analysis of 2420 published articles on the Prisoner's Dilemma between 1951 and
2018. It presents a number of research topics in the PD publications, which have been
identified using Latent Dirichlet Allocation (LDA)~\citep{Blei2003}, and it explores the changes in the
research topics over time. The collaborative behaviour of the field is explored
using the co-authorship network, and furthermore, the LDA
topic analysis is combined with the co-authorship network analysis to assess
the most central authors in these topics. Assessing the collaborative
behaviour of the field of collaboration itself is the main aim of this work.

As discussed in~\citep{youngblood2018}, bibliometrics (the statistical analysis
of published works originally described by~\citep{pritchard1969}) has been used
to support historical assumptions about the development of fields
\citep{raina1998}, identify connections between scientific growth and policy
changes \citep{das2016}, develop a quantitative understanding of author
order~\citep{sekara2018}, and investigate the collaborative structure of an
interdisciplinary field~\citep{Liu2015}. Most academic research is undertaken in
the form of collaborative effort and as~\citep{Kyvik2017} points out, it is
rational that two or more people have the potential to do better as a group
than individually. Indeed this is the very premise of the PD itself.
Collaboration in groups has a long tradition in experimental
sciences and it has be proven to be productive according
to~\citep{Etzkowitz1992}. The number of collaborations can be different
between research fields and understanding how collaborative a field is not
always an easy task. Several studies tend to consider academic citations as a
measure for these things. A blog post published by Nature~\citep{nature_blog}
argues that depending on citations can often be misleading because the true
number of citations can not be known. Citations can be missed due to data entry
errors, academics are influenced by many more papers than they actually cite and
several of the citations are superficial.

A more recent approach to measuring collaborative behaviour, and to studying the
development of a field is to use the co-authorship network, as described
in~\citep{Liu2015}. The co-authorship network has many advantages as several
graph theoretic measures can be used as proxies to explain author relationships.
For example the average degree of a node corresponds to the average number of
an authors' collaborators, and clustering coefficient corresponds to the extent that
two collaborators of an author also collaborate with each other.
In~\citep{Liu2015}, the approach was applied to analyse the development of the field
``evolution of cooperation'', and in~\citep{youngblood2018} to identify the
subdisciplines of the interdisciplinary field of ``cultural evolution'' and
investigate trends in collaboration and productivity between these subdisciplines.
Moreover, \citep{Li2019} examined the
long-term impact of co-authorship with established, highly-cited scientists on
the careers of junior researchers.

LDA is a topic modeling technique proposed
in~\citep{Blei2003} as a generative probabilistic model for discovering
underlying topics in collections of data.
Applications of the technique include detection in image data~\citep{
Coelho2010} and detection in video~\citep{Wang2008}. Nevertheless,
LDA has been applied by several works on publication data for identifying the
topic structure of a subject area. In~\citep{Inglis2018}, it was applied to the
publications on mathematical education of the journals ``Educational Studies in
Mathematics'' and ``Journal for Research in Mathematics Education'' to
identify the dominant topics that each journal was publishing on. The topics of
the North American library and Information Science dissertations were 
studied chronologically in~\citep{Sugimoto2011}, and the main topic of the
scientific content presented at EvoLang conferences was identified
in~\citep{Bergmann2018}. In~\citep{Bergmann2018} the LDA approach is combined with
clustering and a co-authorship network analysis. A clustering analysis is
applied to the LDA topics, and the co-authorship network is analysed as a whole
where the clusters are only used to differentiate between the authors' topics.

This paper builds upon the previous works of~\citep{Bergmann2018, Liu2015,
youngblood2018}. It extends their methodology, it combines identified topics by
an LDA model with the co-authorship network analysis, and applies all these
techniques to a new data set. This data set was collected not from a single
source but from five different sources. The four publishers were chosen because
they are well known publishers in the field, and the arXiv preprint server. 
The search
terms used to collect data appear on relevant articles and the search fields
that were used were the title, abstract and text. However, papers can refer to
the PD in the text but not analyze the topic. For this reason such articles were
manually checked, so that only relevant papers are included in the analysis.
Moreover, an amount of well
known articles, which are not published in any of the selected publishers, were
manually included in the data set.

The methodology used in this manuscript, which
includes the data collection and a preliminary analysis of the data set, is
covered in Section~\ref{section:methodology}. The results on the research topics
of the PD are presented in Section~\ref{subsection:research_topics}, and the
results on the co-authorship network are presented in
Section~\ref{subsection:co_authorship}. Finally, the conclusions are summarised
in Section~\ref{section:conclusion}.

\section{Methodology}\label{section:methodology}

Academic articles are accessible through scholarly databases. Several databases
and collections today offer access through an open application protocol
interface (API). An API allows users to query directly a publisher's database and
bypass the graphical user interface. Interacting with an API has two phases:
requesting and receiving. The request phase includes composing a url with the
details of the request. For example,
\url{http://export.arxiv.org/api/query?search_query=abs:prisoner's
dilemma&max_results=1} represents a request message. The first part of the
request is the address of the API. In this example the address corresponds to
the API of arXiv. The second part of the request contains the search arguments.
In this example it is requested that the word `prisoners dilemma' exists within
the article's title. The format of the request message is different from API to
API. The receive phase includes receiving a number of raw metadata of articles
that satisfies the request message. The raw metadata are commonly received in
extensive markup language (xml) or Javascript object notation (json)
formats~\citep{nurseitov2009}. Similarly to the request message, the structure of
the received data differs from publisher
to publisher.

The data collection is crucial to this study. To ensure that this study can be
reproduced all code used to query the different publishers' APIs has been packaged as a
Python library and is available online~\citep{nikoleta_2017}. The software could
be used for any type of projects similar to the one described here,
documentation for it is available at:
\url{http://arcas.readthedocs.io/en/latest/}. Project~\citep{nikoleta_2017} can
collect data from five different sources. These correspond to four
publishers and a preprint server:

\begin{multicols}{2}
    \begin{itemize}
        \item arXiv~\citep{mckiernan2000}; a repository of electronic preprints.
        It consists of scientific
        papers in the fields of mathematics, physics, astronomy, electrical engineering,
        computer science, quantitative biology, statistics, and quantitative finance,
        which all can be accessed online.
        \item PLOS~\citep{plos}; a library of open access journals and other scientific literature
        under an open content license. It launched its first journal, PLOS Biology,
        in October 2003 and publishes seven journals, as of October 2015.
        \item IEEE Xplore Digital Library (IEEE)~\citep{ieee}; a research database for discovery
        and access to journal articles, conference proceedings, technical standards,
        and related materials on computer science, electrical engineering and electronics,
        and allied fields. It contains material published mainly by the Institute of
        Electrical and Electronics Engineers and other partner publishers. 
        \item Nature~\citep{nature}; a multidisciplinary scientific journal,
        first published on 4 November 1869. It was ranked the world's most cited
        scientific journal by the Science Edition of the 2010 Journal Citation Reports
        and is ascribed an impact factor of 40.137, making it one of the world's
        top academic journals.
        \item Springer~\citep{springer}; a leading global scientific publisher of
        books and journals. It publishes close to 500 academic and professional
        society journals.
    \end{itemize}
\end{multicols}

These publishers were chosen because they are prominent publishers in the field. For
each source, data can be collected by specifying a search term and a search field.
Articles for which any of the terms:

\begin{itemize}
    \item prisoner's dilemma
    \item prisoners dilemma
    \item prisoner dilemma
    \item prisoners evolution
    \item prisoner game theory
\end{itemize}

existed within the title, the abstract or the text are included in the analysis.
These terms we selected because they are occurring terms in paper known to be
relevant in the field. However, the authors acknowledge that there are other
terms that could have been used, for example ``donation game''. The authors believe
that the results of the manuscript do generalise to the overall stated goals
(Section~\ref{section:introduction}), but they are inferred only from
the data collected on the specific search terms and search fields.

The latest data collection was performed on the \(30^{\text{th}}\) November
2018. Following the automatic collection of articles from the sources, a
cleaning process was applied to the data. More specifically, all the titles of
the collected articles were compared for semantic similarity. There were a total
of 34 duplicate articles. That was because both the preprint and the published
versions of a paper were collected. The preprint versions
(collected from arXiv) were dropped at this stage. A semantic similarity check
was also applied in the names of the collected authors. The names that were
highlighted as similar were manually checked. In case of a duplicate, for
example ``Martin Nowak'' and ``Martin A. Nowak'' are considered duplicates, all
entries of that author were fixed to a single style. Most commonly the middle
name was dropped. Finally, articles that were collected because the search terms
existed within the text were checked to reassure their relevance to the PD
topic. Non relevant articles were dropped at this stage. 

\begin{figure}[!hbtp]
    \begin{center}
        \includegraphics[width=.8\textwidth]{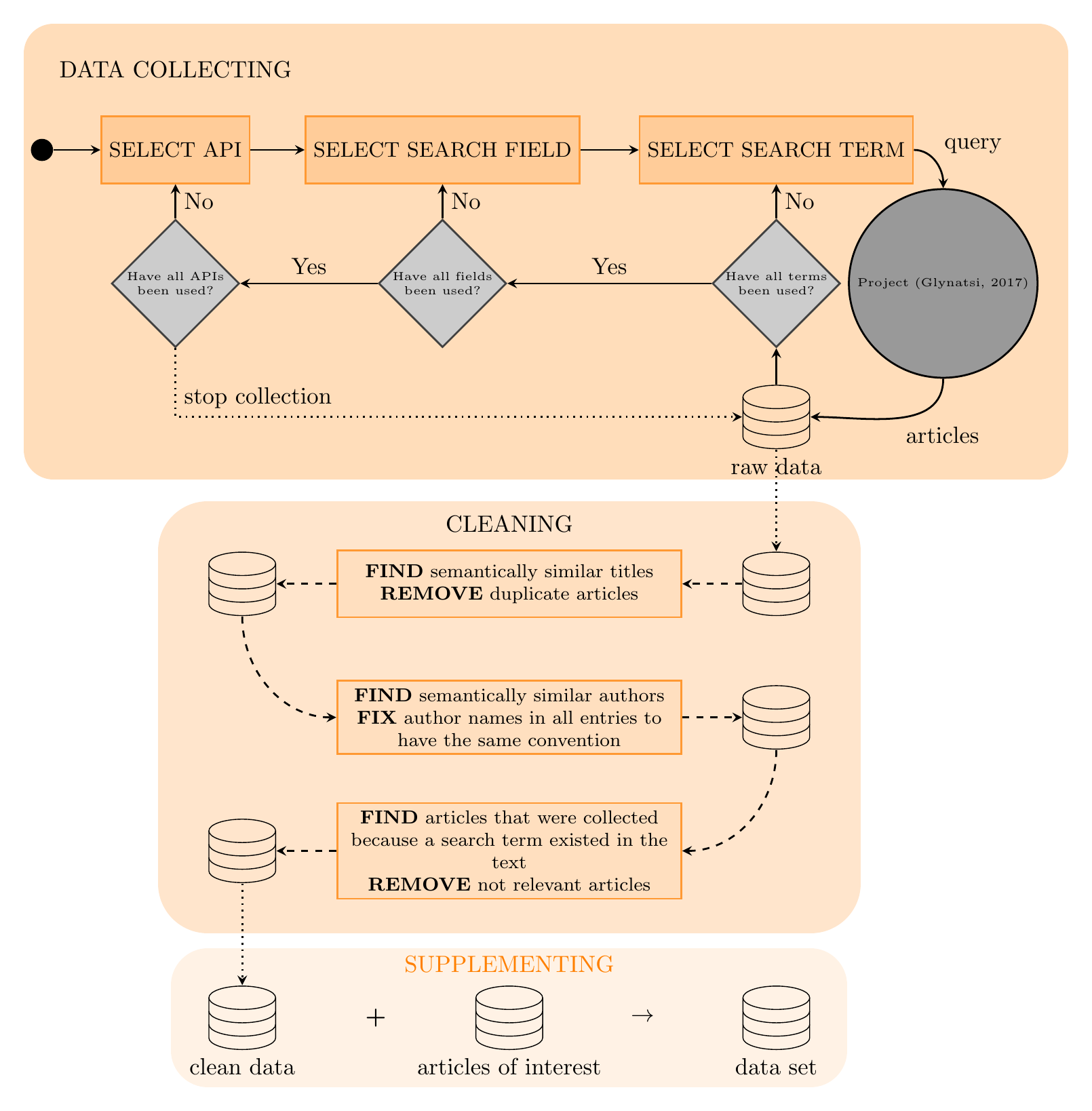}
    \end{center}
    \caption{The generating process of the data set~\citep{pd_data_2018}.}\label{figure:colection_process}
\end{figure}

Following the cleaning process, a total of \manual~articles were manually added
to the data set because they are of interest to the field. This was also done
in~\citep{Liu2015}. Examples of such papers include~\citep{Flood1958} the first publication on the
PD,~\citep{Ohtsuki2006, Stewart2012} two well cited articles in the field, and a
series of works from Robert Axelrod~\citep{Axelrod1980, Axelrod1980more,
Axelrod1987, Axelrod1981, Riolo2001} a leading author of the field.
The process of obtaining the data set used in analysis presented in the
manuscript is illustrated in Figure~\ref{figure:colection_process}.
This data set has been archived and is available at~\citep{pd_data_2018}.

The data set consists of \totalarticles~articles with unique
titles. A more detailed summary of the articles' provenance
is given by Table~\ref{table:preliminary_table}. Only 3\% of the data set consists of
articles that were manually added and 27\% of the articles were collected from
arXiv. The average number of publications is also included in
Table~\ref{table:preliminary_table}. Overall an average of 43 articles are published
per year on the topic. The most significant contribution to this appears to be
from arXiv with 11 articles per year, followed by Springer with 9 and PLOS with
8.

\begin{table}[!hbtp]
    \begin{center}
    \resizebox{.9\textwidth}{!}{
        \begin{tabular}{lrrrr}
            \toprule
            {} &  Number of Articles &  Percentage \% &  Year of first publication &  Average number of publications per year\\
            \midrule
            IEEE     &               294 &       12.14\% &                    1973 &                             5\\
            Manual   &                76 &        3.14\% &                    1951 &                             1\\
            Nature   &               436 &       18.00\% &                    1959 &                             8\\
            PLOS     &               477 &       19.69\% &                    2005 &                             8\\
            Springer &               533 &       22.01\% &                    1966 &                             9\\
            arXiv    &               654 &       27.00\% &                    1993 &                            11\\
            Overall  &              2470 &      100.00\% &                    1951 &                            43\\
            \bottomrule
        \end{tabular}
    }
    \end{center}
    \caption{Summary of~\citep{pd_data_2018} per provenance.}
    \label{table:preliminary_table}
\end{table}

All the visualisations presented in the manuscript were generated
using~\citep{hunter2007matplotlib}, and project~\citep{walt2011numpy} was used for
manipulating the data.

\begin{figure}[!hbtp]
    \centering
    \includegraphics[width=.50\textwidth]{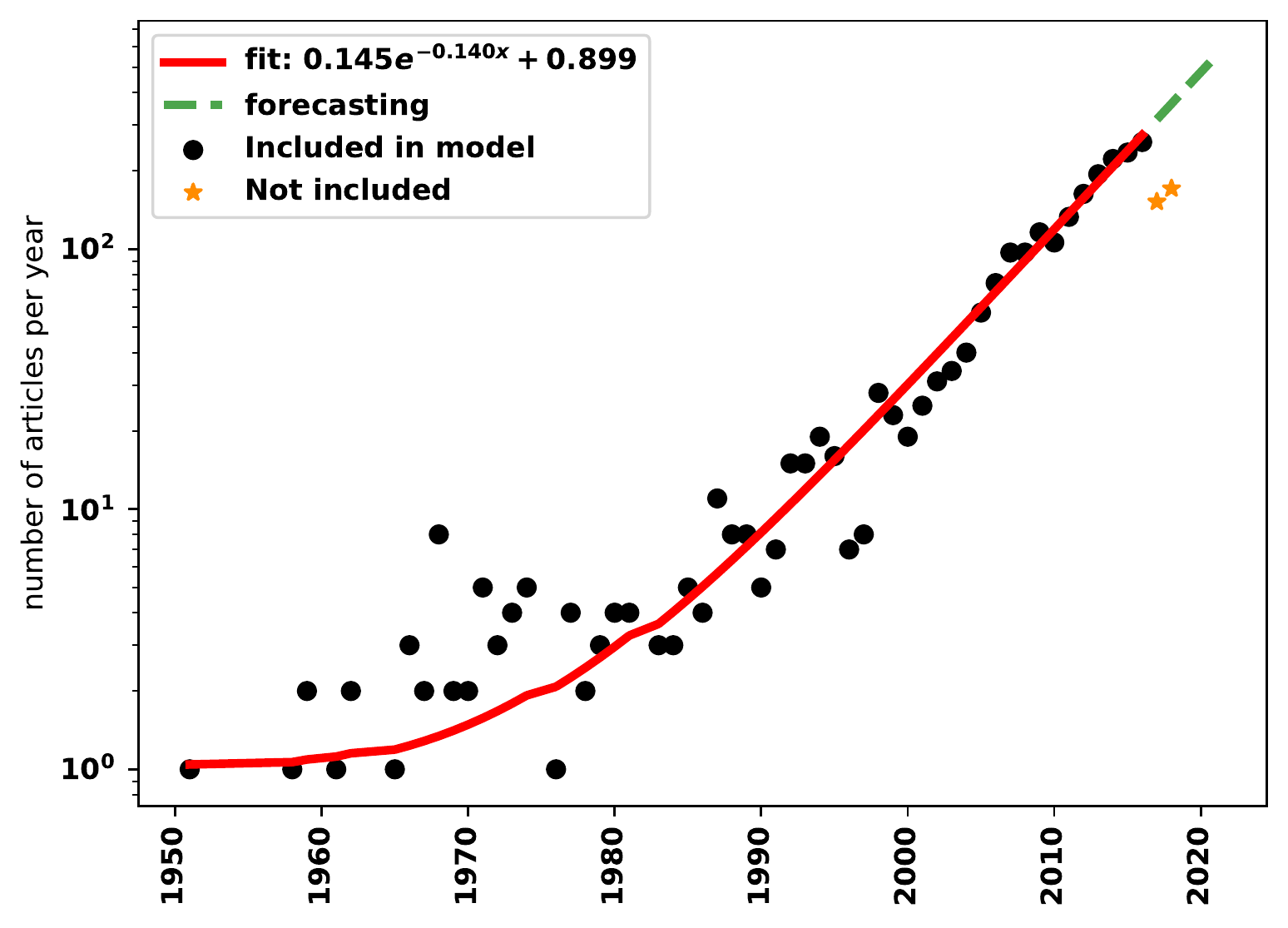}
    \caption{Number of articles published on the PD 1951-2018 (on a log scale),
    with a fitted exponential line, and a forecast for 2017-2022.}\label{fig:timeseries}
\end{figure}

The data handled here is in fact a time series from the 1950s, the formulation
of the game, until 2018 (Figure~\ref{fig:timeseries}). Two observations can be
made from Figure~\ref{fig:timeseries}.

\begin{enumerate}
    \item There is a steady increase of the number of publications since the
    1980s and the introduction of computer tournaments~\citep{Axelrod1981}
    (work by Robert Axelrod).
    \item There is a decrease in 2017-2018. This is due to our data set being
    incomplete. Articles that have been written in 2017-2018 have either not
    being published or were not retrievable by the APIs at the time of the last
    data collection.
\end{enumerate}

These observations can be confirmed by studying the time series.
Using~\citep{scipy}, an exponential distribution is fitted to the data.
The fitted model can be used to forecast the
behaviour of the field for the next 5 years. Even
though the time series has indicated a slight decrease, the model forecasts that
the number of publications will keep increasing, thus demonstrating that the
field of the PD continues to attract academic attention.

There are a total of \authors~authors in the data set \citep{pd_data_2018} and several of these
authors have had multiple publications collected from the data collection process.
The highest number of articles collected for an
author is 83 publications for Matjaz Perc. However, Matjaz Perc is an outlier
most authors have 1 to 6 publications in the data set.
The overall Collaboration Index (CI) or the average number of authors on
multi-authored papers is 3.2, thus on average a non single author publication in
the PD has 3 authors. This appears to be quite standard compared to other fields
such as cultural evolution~\citep{youngblood2018}, Astronomy and Astrophysics,
Genetics and Heredity, Nuclear and Particle Physics as reported
by~\citep{nature_author_blog}.
There are only a total of 545 publications with a single author, which
corresponds to the 22\% of the papers. It appears that academic publications
tend to be undertaken in the form of collaborative effort, which is in line
with the claim of~\citep{Kyvik2017}.

The collaborativeness of the authors is explored in more detail in
Section~\ref{subsection:co_authorship} using the co-authorship network. The collaborative
behaviour of authors will also be explored at the research topics level. These topics and their
relevance over time are presented in Section~\ref{subsection:research_topics}.

\section{Results}\label{section:results}

\subsection{Research topics in the Prisoner's Dilemma research}\label{subsection:research_topics}

The articles contained in the data set \citep{pd_data_2018} are classified
into research topics using LDA, an unsupervised machine learning technique
designed to summarize large collections of documents by a small number of
conceptually connected topics or themes~\citep{Blei2003, Grimmer2013}. The
documents are the articles' abstracts and LDA was carried out using~\citep{rehurek_lrec}.
In LDA, each document/abstract is represented by a distribution over topics,
and the topics themselves are represented by a distribution over words. More
specifically, each topic is described by weights associated with words and
each document by the probabilities of belonging to a specific topic. The
probability of a document belonging to topic \(T\) is referred to as the percentage
contribution denoted as \(c_T\). For example the words and their associated
weights for two topics A and B could be:

\begin{itemize}
    \item Topic A: \(0.039 \times\)``cooperation'', \(0.028 \times\)``study'' and \(0.026 \times\)``human''.
    \item Topic B: \(0.020 \times\)``cooperation'', \(0.028 \times\)``agents'' and
    \(0.026 \times\)``strategies''.
\end{itemize}

The percentage contribution for a document with abstract ``The study of
cooperation in humans'' has a \(c_{A} = 0.039 + 0.028 + 0.026 = 0.093\) and
\(c_B = .020 + 0.0 + 0.0 = 0.020\). The topic to which a document is assigned to
is based on the highest percentage contribution denoted as \(c^*\). For the
given example the dominant topic is Topic A \(c^*=c_A\).

\begin{figure}[!hbtp]
    \centering
    \begin{subfigure}{.45\textwidth}
        \centering
        \includegraphics[width=\textwidth]{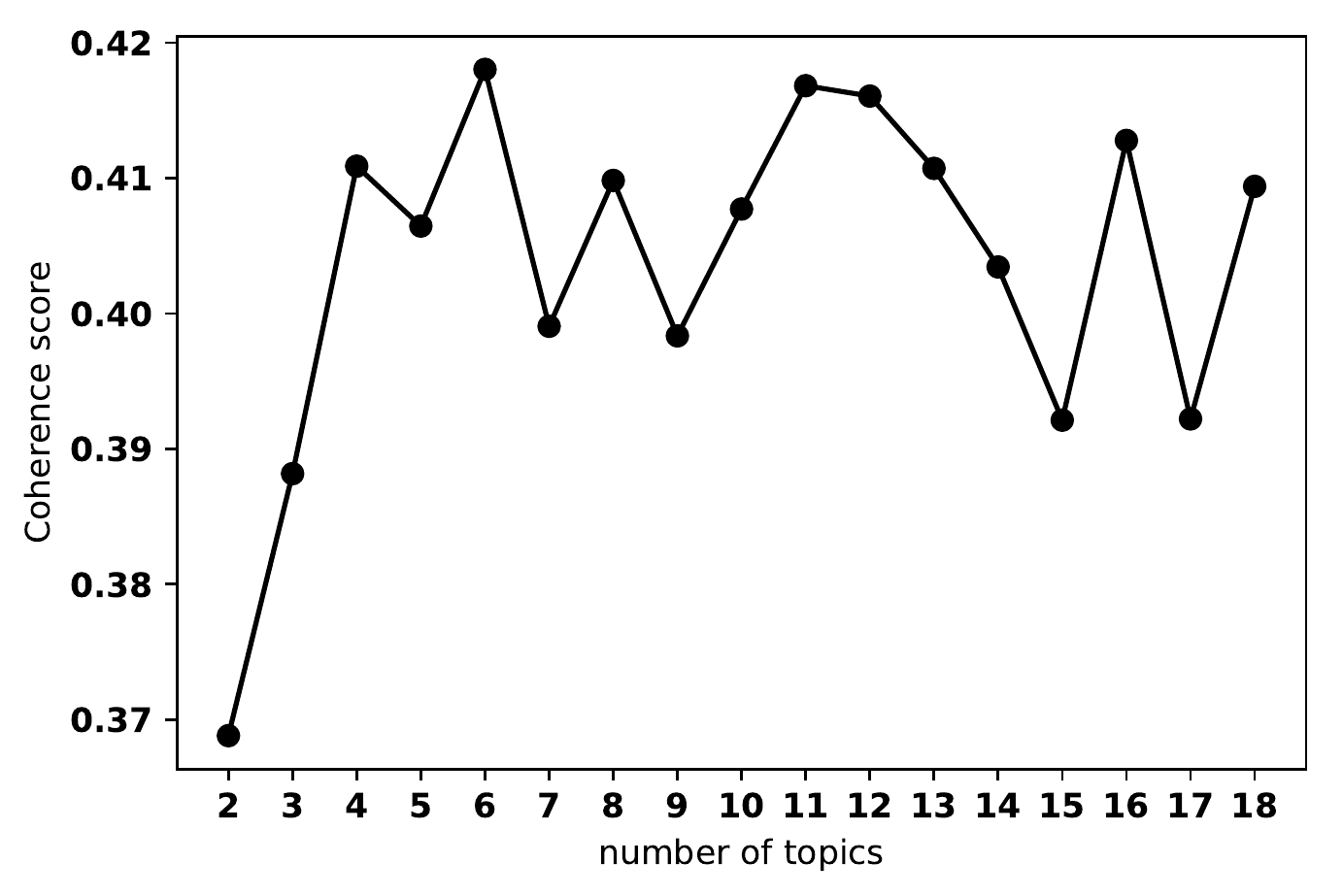}
        \caption{Coherence for LDA models over the number of topics.}
        \label{fig:coherence_value_over_number_of_topics}
    \end{subfigure}\hfill
    \begin{subfigure}{.45\textwidth}
        \centering
        \includegraphics[width=\textwidth]{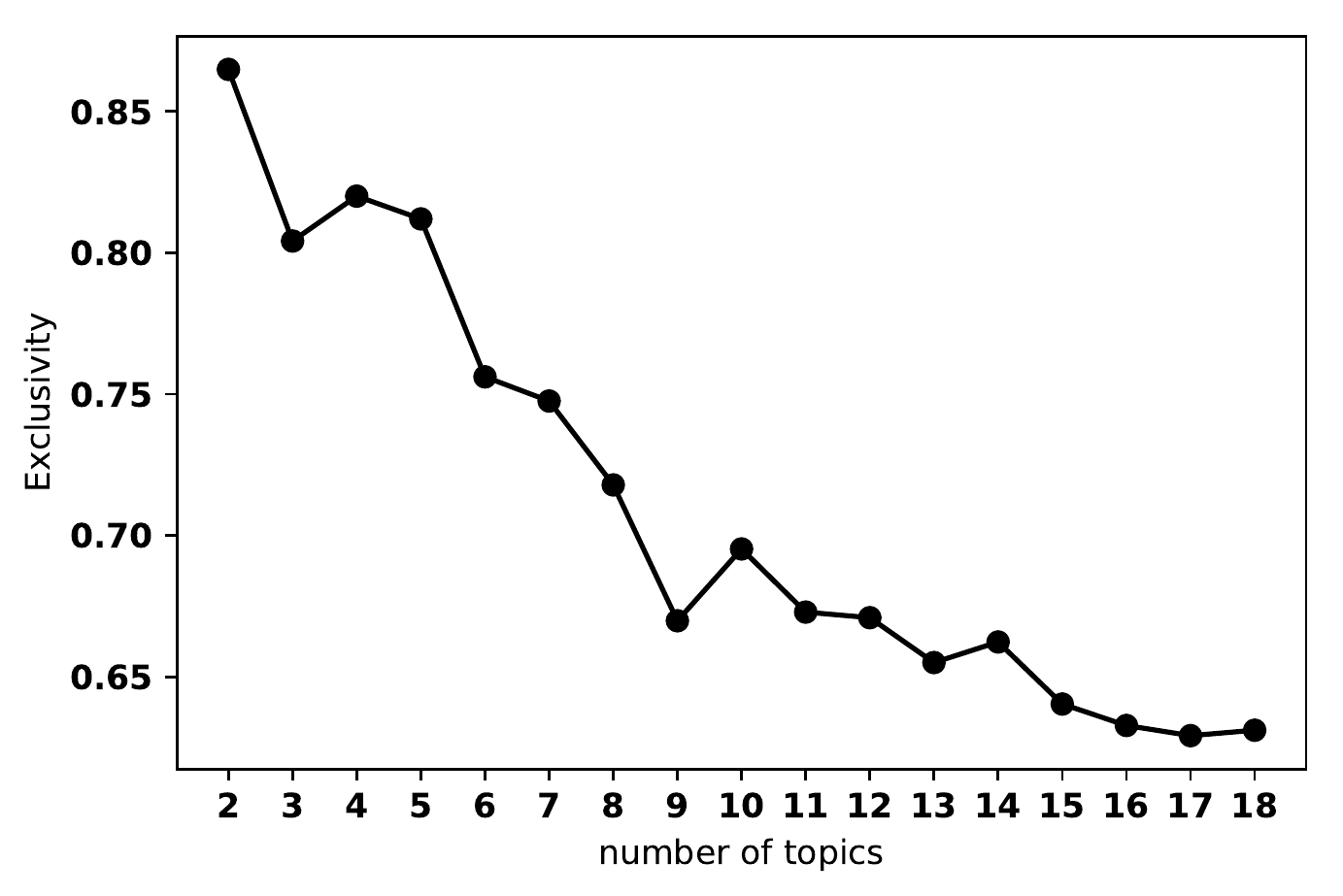}
        \caption{Exclusivity for LDA models over the number of topics.}
        \label{fig:exclusivity_value_over_number_of_topics}
    \end{subfigure}
    \caption{Coherence and exclusivity for LDA models over the number of topics.}\label{figure:models_scores}
\end{figure}

LAD requires that the number of topics is specified in advance before running
the algorithm. The appropriate number of topics can be chosen based on the
coherence score~\citep{Roder2015} or the exclusivity score~\citep{Airoldi2012}. The
coherence score measures the degree of semantic similarity between highly weighted
words of a topic. There are cases for which a few topics can be dominated by very
common words, and for that reason the exclusivity of words to topics has also
been calculated.
Figure~\ref{fig:coherence_value_over_number_of_topics} gives the topic coherence
and Figure~\ref{fig:exclusivity_value_over_number_of_topics} gives the
exclusivity of 18 models where the number of topics \(n \in \{2, 3, \dots, 18\}\).
The topic coherence for each model was calculated using the
open source project~\citep{rehurek_lrec}. The exclusivity measure was calculated
with an altered version of~\citep{rehurek_lrec} which has been archived
at~\citep{gensim_nikoleta}.

From Figure~\ref{fig:coherence_value_over_number_of_topics} it can be seen that
the number of topics with the highest coherence score are $n=6$ (coherence score
of 0.418) and $n=12$ (coherence score of 0.417).
Figure~\ref{fig:coherence_value_over_number_of_topics} shows that the
exclusivity of the highly weighted words of the topics is decreasing as the number
of topics increases. A number of topics $n=5$ has a better exclusivity value
than the model of $n=6$, and its coherence score is 0.406 (which is closed to
0.418). For that reason $n=5$ is chosen to carry out the analysis of this work.

For \(n=5\) the articles are clustered and assigned to their dominant topic,
based on the highest percentage contribution. The keywords associated with a
topic, the most representative article of the topic (based on the
percentage contribution) and its academic reference are given by
Table~\ref{table:topics_and_articles}. The topics are labelled as A, B, C, D and
E, and more specifically:

\begin{itemize}
    \item Based on the keywords associated with Topic A, and the most
    representative article, Topic A appears to be about \textbf{human subject
    research}. Several publications assigned to the topic study the PD by
    setting experiments and having human participants simulate the game
    instead of computer simulations. These articles include~\citep{Matsumoto2016}
    which showed that prosocial behavior increased with the age of the
    participants, ~\citep{Li2014} which studied the difference in cooperation
    between high-functioning autistic and typically developing
    children,~\citep{Molina2013} explored the gender effect in highschool
    students and~\citep{Bell2017} explored the effect of facial expressions of
    individuals.
    \item Though it is not immediate from the keywords associated with
    Topic B, investigating the papers assigned to the topic indicate that it
    is focused on \textbf{biological studies}. Papers assigned to the topic include
    papers which apply the PD to genetics~\citep{Sistrom2015}, to
    the study of tumours~\citep{sartakhti2017} and
    viruses~\citep{Turner1999}. Other works include how phenotype affinity
    can affect the emergence of cooperation~\citep{wu2019phenotype} and modeling
    bacterial communities as a spatial structured social dilemma.
    \item Based on the keywords and the most representative article Topic
    C appears to include publications on PD \textbf{strategies}. Publications
    in the topic include the introduction of new strategies~\citep{stewart2013extortion},
    the search of optimality in strategies~\citep{banerjee2007reaching} and the
    training of strategies~\citep{ishibuchi2011evolution} with different
    representation methods. Moreover, publications that study the evolutionary
    stability of strategies~\citep{adami2013evolutionary} and introduced methods
    of differentiating between them~\citep{ashlock2008fingerprinting} are
    also assigned to C.
    \item The keywords associated with Topic D clearly show that the topic
    is focused on \textbf{evolutionary dynamics on networks}. Publications include
    \citep{ichinose2013robustness} which explored the robustness of cooperation
    on networks,~\citep{wang2012spatial} which studied the effect of a strategy's neighbourhood
    on the emergence of cooperation and~\citep{chen2016fixation} which explored
    the fixation probabilities of any two strategies is spatial
    structures.
    \item The publication assigned to Topic E are on \textbf{modeling problems
    as a PD game}. Though Topic B is also concerned with problems being formulated
    as a PD, it includes only biological problems. In comparison, the problems
    in Topic E include decision making in
    operational research~\citep{ormerod2010or}, information sharing among members
    in a virtual team~\citep{feng2008trilateral}, the measurement of influence
    in articles based on citations~\citep{hutchins2016relative} and the price
    spikes in electric power markets~\citep{Guan2002}, and not on biological studies.
\end{itemize}

\begin{table}[!hbtp]
    \begin{center}
    \resizebox{\textwidth}{!}{
    \begin{tabularx}{1.5\textwidth}{lXXl|cc}
        \toprule
        Dominant Topic &                                                                                                 Topic Keywords &                                                                                                                                    Most Representative Article Title &        Reference &  \# Documents &  \% Documents \\
        \midrule
        A &                 social, behavior, human, study, experiment, cooperative, cooperation, suggest, find, behaviour &                                                                                      Facing Aggression: Cues Differ for Female versus Male Faces &  \citep{Geniole2012} &                496.0 &                   0.2008 \\
        B &                               individual, group, good, show, high, increase, punishment, cost, result, benefit &  Genomic and Gene-Expression Comparisons among Phage-Resistant Type-IV Pilus Mutants of Pseudomonas syringae pathovar phaseolicola &  \citep{Sistrom2015} &                309.0 &                   0.1251 \\
        C &                             game, strategy, player, agent, dilemma, play, payoff, state, prisoner, equilibrium &                                                            Fingerprinting: Visualization and Automatic Analysis of Prisoner's Dilemma Strategies &  \citep{ashlock2008fingerprinting} &                561.0 &                   0.2271 \\
        D &  cooperation, network, population, evolutionary, evolution, interaction, dynamic, structure, cooperator, study &                                                   Influence of initial distributions on robust cooperation in evolutionary  Prisoner's Dilemma &     \citep{Chen2007} &                556.0 &                   0.2251 \\
        E &                           model, theory, base, system, problem, paper, propose, information, provide, approach &                                                                          Gaming and price spikes in electric power markets and possible remedies &     \citep{Guan2002} &                548.0 &                   0.2219 \\
        \bottomrule
    \end{tabularx}
    }
    \end{center}
    \caption{Keywords for each topic and the document with the most representative article for each topic.}
    \label{table:topics_and_articles}
\end{table}

Note that the whilst for the choice of 5 topics the actual clustering is not
subjective (the algorithm is determining the output) the interpretation above is.

\begin{figure}[!hbtp]
    \centering
    \includegraphics[width=\textwidth]{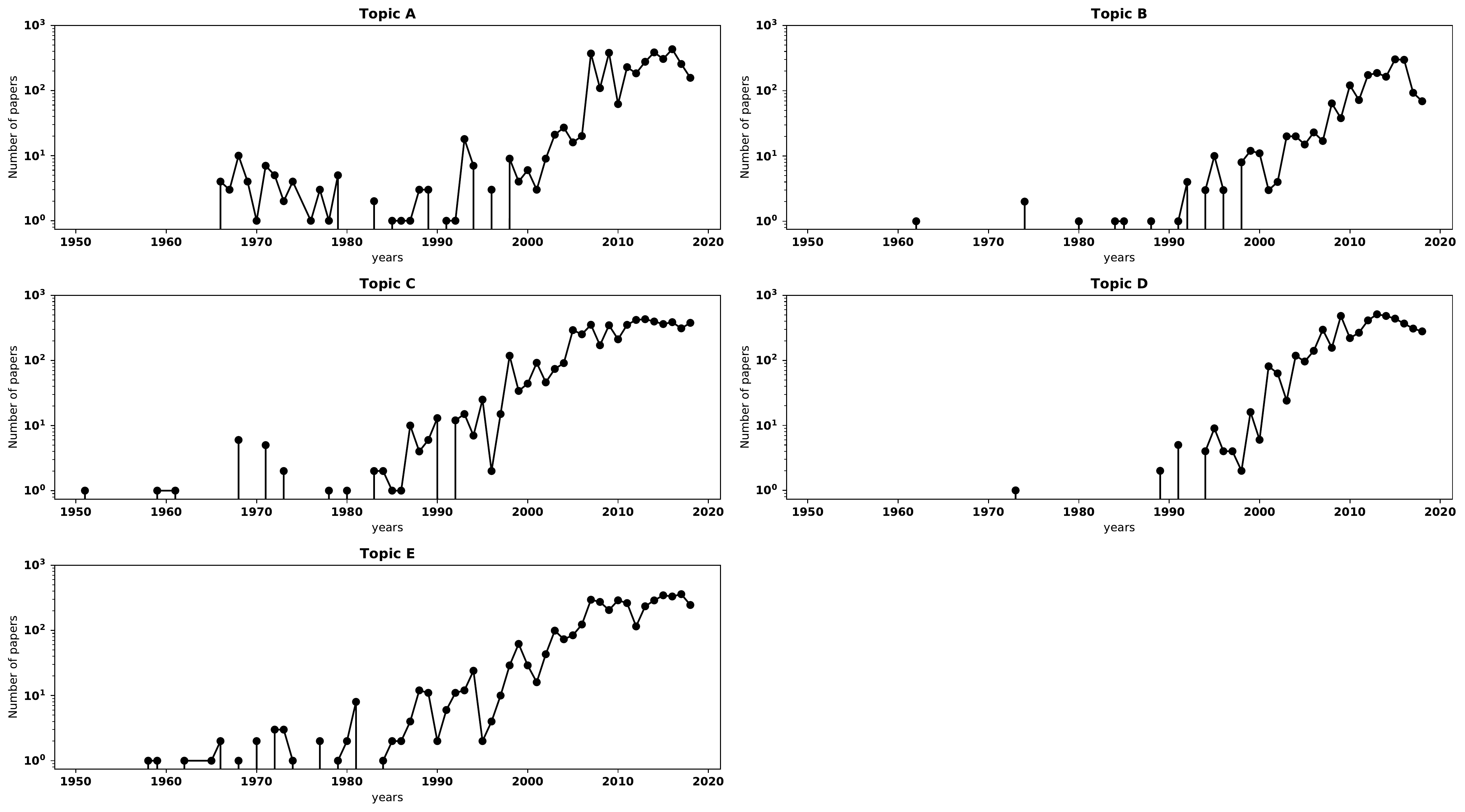}
    \caption{Number of articles per topic over the years (on a logged scale).}\label{fig:number_of_articles_per_topic}
\end{figure}

Figure~\ref{fig:number_of_articles_per_topic} gives the number of articles
per topic over time. The topics appear to have had a similar trend over the years,
with topics B and D having a later start. Following the introduction of a topic
the publications in that topic have been increasing, and there is no decreasing
trend in any of the topics. All the topics have been publishing for years and
they still attract the interest of academics. Thus, there does not
seem to be any given topic more or less in fashion.

To gain a better understanding regarding the change in the topics over the years,
LDA is applied to the cumulative data set over 8 time periods. These periods are
1951-1965, 1951-1973, 1951-1980, 1951-1988, 1951-1995, 1951-2003, 1951-2010,
1951-2018. The number of topics for each cumulative subset is chosen based only
on the topic coherence, and the exclusivity is not taken into account. As a result,
the period 1951-2018 has been assigned \(n=6\) which had the highest coherence
value instead of 5. The chosen models for each period including the
number of topics, their keywords and number of articles assigned to them are
given in the Appendix~\ref{appendix:topics_per_year}.

But how well do the five topics which were presented earlier fit the
publications over time? This is answered by comparing the performance of three
LDA models over the cumulative periods' publications. The three models are LDA
models for the entire data set for \(n\) equal to 5, 6 and the models of
Table~\ref{table:topics_per_year} for each time period. Thus, for the period
1951-1980 the three model that are being compared are for $n$ equal to 5, 6, and
13.

\begin{figure}[!hbtp]
    \centering
    \includegraphics[width=.75\textwidth]{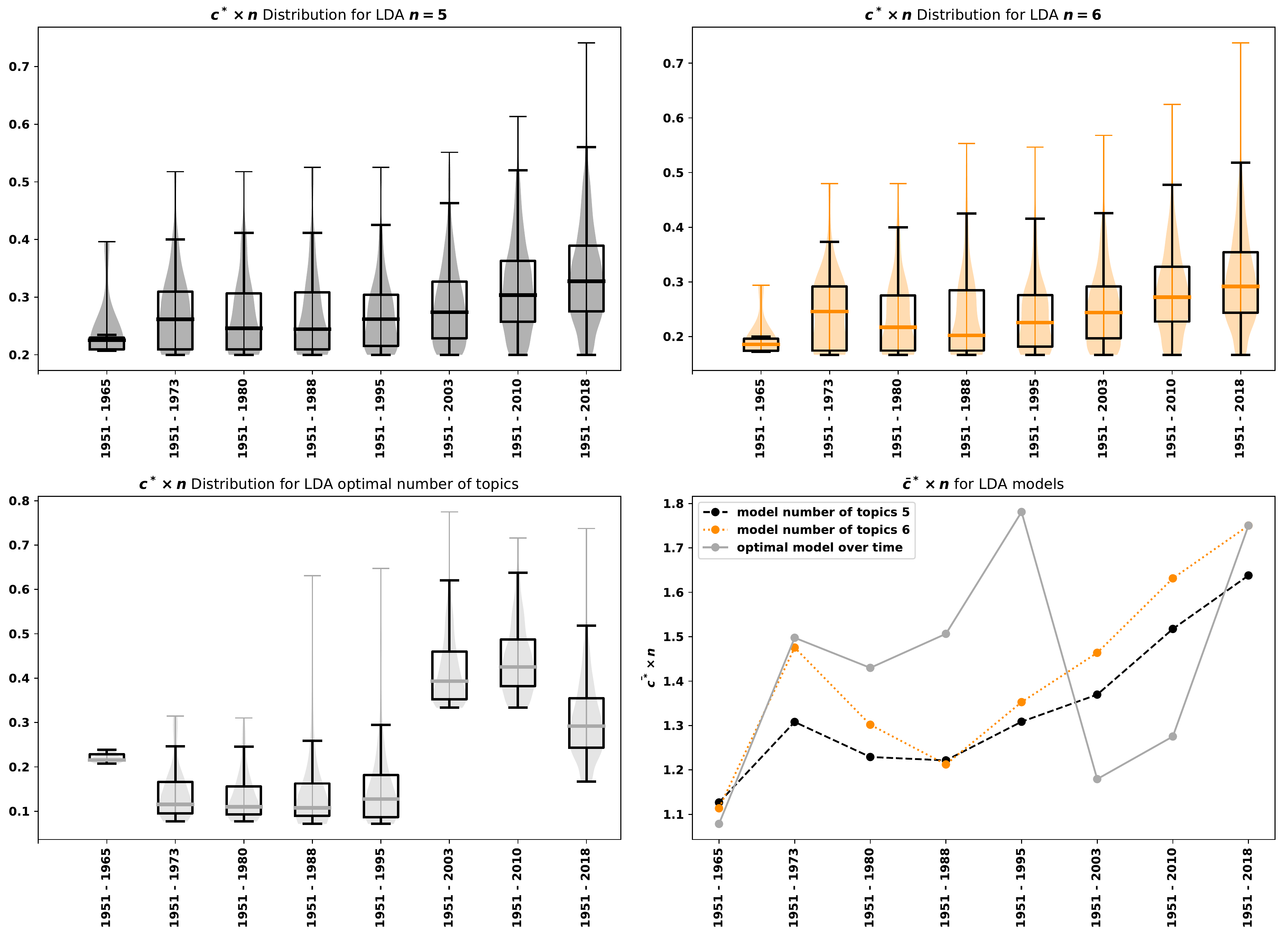}
    \caption{Maximum percentage contributions (\(c^*\)) over the time periods,
    for the LDA models for the entire data set for \(n\) equal to 5, 6
    and for the models of Table~\ref{table:topics_per_year}.}
    \label{fig:median_percentage_contribution_over_time}
\end{figure}

For each model the \(c^*\) is estimated for each document in the cumulative data
sets. The performance of the models are then compared based on:

\begin{equation}\label{eq:ratio}
    \bar{c^*} \times n
\end{equation}

where \(\bar{c^*}\) is the median highest percentage contribution and \(n\)
is the number of topics of a given period. A model with more topics will have more
difficulty to assign papers. Thus, equation (\ref{eq:ratio}) is a measure of confidence
in assigning a given paper to its topic weighted by the number of topics.
The performances are
given by Figure~\ref{fig:median_percentage_contribution_over_time}.

The five topics of the PD presented in this manuscript appear to always be
less good at fitting the publications compared to the six topics of LDA \(n=6\).
Moreover, these are less good than the models of periods 1951-1965 to 1951-1995.
The difference in the performance values, equation (\ref{eq:ratio}),
however is small. The relevances of the five topics has been increasing
over time, and though, the topics did not always fit the majority of published
work, there were still papers being published on those topics.

In the following section the collaborative behaviour of authors in the field,
and within the field's topics as were presented in this section, are explored
using a network theoretic approach.

\subsection{Analysis of co-authorship network}\label{subsection:co_authorship}

The relationship between the authors within a field is modeled as a graph
\(G = (V_G, E_G)\) where \(V_G\) is the set of nodes and \(E_G\)  is the set of
edges. The set \(V_G\) represents the authors and an edge connects two authors
if and only if those authors have written together. This co-authorship network is
constructed using the main data set~\citep{pd_data_2018} and the open source package
\citep{networkx}. The PD network is denoted as \(G\) where the
number of unique authors \(|V(G)|\) is \authors~and \(|E(G)|\) is \edges.

The collaborativeness of the authors is analysed using measures such as, number of isolated nodes,
number of connected components, clustering coefficient, number of communities, modularity and average degree.
These measures show the number of connections authors can have
and how strongly connected these people are. The number of isolated nodes is the
number of nodes that are not connected to another node, thus the
number of authors that have published alone. The average degree denotes the average
number of neighbours for each nodes, i.e. the average number of collaborations
between the authors.
A connected component is a maximal set of nodes such that each pair of nodes is
connected by a path~\citep{Easley2010}. The number of connected components as well as the size of the
largest connected component in the network are reported.
The size of the largest connected component represents the scale of the central cluster
of the entire network.
Clustering coefficient and modularity are also calculated. The clustering
coefficient, defined as 3 times the number of triangles on the graph divided
by the number of connected triples of nodes, is a local measure of the degree to
which nodes in a graph tend to cluster together
in a clique~\citep{Easley2010}. It shows to which extent the collaborators
of an author also write together.
In comparison, modularity is a global measure designed to measure the strength of
division of a network into communities. The number of communities is reported
using the Clauset-Newman-Moore method~\citep{clauset2004}. Also the modularity index
based on the Louvain method~\citep{Blondel2008} is calculated using~\citep{python_louvain}. The value
of the modularity index can vary between \([-1, 1]\), a high value of modularity
corresponds to a structure where there are dense connections between the nodes within
communities but sparse connections between nodes in different communities.
That means that there are many sub communities of authors that write together
but not across communities.
Two centrality measures are also reported. These are:

\begin{enumerate}
    \item \textbf{Closeness centrality}, where a node
    is seen as centrally involved in the network if it requires only few
    intermediaries to contact others and thus is structurally relatively
    independent.
    \item \textbf{Betweenness centrality},
    where the determination of an author's centrality is based on the quotient
    of the number of all shortest paths between nodes in the network that
    include the node in question and the number of all shortest paths in the
    network. In betweenness centrality the position of the node matters.
\end{enumerate}

There are a total of
\connectedcomponents~connected components in \(G\) and the largest component has
a size of \largestcc~nodes. The largest connected component is going to be
refereed to as the main cluster of the network and is denoted as \(\bar{G}\). A
metrics summary of both networks is given
by Table~\ref{table:network_comparison}.
Based on Table~\ref{table:network_comparison} an author in \(G\) has on
average 4 collaborators and a 67\% probability of collaborating with a
collaborator's co-author. An author of \(\bar{G}\) on average is 10\% more likely
to write with a collaborator's co-author and on average has 2 more
collaborators. Moreover, there are only \isolatedpercentage\% of authors in the
PD that have no connection to any other author.

How does this compare to other fields? Two more data sets for the topics
``Price of Anarchy'' and ``Auction Games'' have been collected in order to
compare the collaborative behaviour of the PD to other game theoretic fields. A
total of 3444 publications have been collected for Auction games and 748 for
Price of Anarchy. Price of Anarchy is relatively a new field, with the first
publication on the topic being~\citep{Koutsoupias1999} in 1999. This explains the
small number of articles that have been retrieved. Both data sets have been
archived and are available in~\citep{auction_data_2018, anarchy_data_2018}.
The networks for both data sets have been generated in the same way as \(G\),
and a summary of the networks' metrics is also given by Table~\ref{table:network_comparison}.

The average degrees for the Price of Anarchy and for Auction games are lower
than the PD's, and so are their respective clustering coefficients. Moreover,
both the Price of Anarchy and Auction games have a larger number of isolated
authors. These results seem to indicate that the PD is a relatively collaborative
field, compared to other game theoretic fields.
However, both \(G\) and \(\bar{G}\) have a high modularity (larger than 0.84) and a large number of
communities (967 and 25 respectively). A high modularity implies that authors create their own publishing
communities but not many publications from authors from different communities
occur. Thus, author tends to collaborate with authors in their communities but
not many efforts are made to create new connections to other communities and
spread the knowledge of the field across academic teams. The fields
of both Price of Anarchy and Auction games also have high modularity, and
that could indicate that is in fact how academic publications are.

\begin{table}[!hbtp]
    \centering
    \resizebox{\textwidth}{!}{
        \begin{tabular}{lrrrrrrrrrr}
            \toprule
            {} &  \# Nodes &  \# Edges &  \# Isolated nodes &  \% Isolated nodes &  \# Connected components &  Size of largest component &  Av. degree &  \# Communities &  Modularity &  Clustering coeff \\
            \midrule
            $G$              &     4221 &     7642 &               338 &               8.0 &                    1157 &                        796 &       3.621 &           1177 &    0.965264 &             0.666 \\
            $\bar{G}$        &      796 &     2214 &                 0 &               0.0 &                       1 &                        796 &       5.563 &             29 &    0.840138 &             0.773 \\
            Auction Games    &     5362 &     7861 &               453 &               8.4 &                    1469 &                       1348 &       2.932 &           1493 &    0.957238 &             0.599 \\
            Price of Anarchy &     1315 &     1952 &               165 &              12.5 &                     406 &                        221 &       2.969 &            414 &    0.964498 &             0.626 \\
            \bottomrule
    \end{tabular}
    }
    \caption{Network metrics for \(G\), \(\bar{G}\), Auction games and Price of
    Anarchy.}
    \label{table:network_comparison}
\end{table}

The evolution of the networks was also explored over time by constructing the
network cumulatively over 51 periods. Except from the first period 1951-1966 the
rest of the periods have a yearly interval (data for the years 1975 and 1982
were not retrieved by the collection data process). The metrics of each sub
network are given in the Appendix~\ref{appendix:tables}.
The results, similarly to the results of~\citep{Liu2015}, confirm that the
networks grow over time and that the networks always had a high modularity.
Since the first publications authors tend to write with people from their
communities, and that is not an effect of a specific time period.

The networks corresponding to the topics of Section~\ref{subsection:research_topics} have
also been generated similarly to \(G\). Note that authors with publications in
more than one topic exist, and these authors are included in all the corresponding
networks. A metrics' summary for all five topic networks is given by Table
\ref{table:topics_networks}.

Topics A and B have the highest average degree and clustering coefficient.
Moreover, both topics have a small number of isolated nodes. Compared to that
Topic C has a smallest average degree and Topic E has the highest number of
isolated authors. These indicate that the topics ``human subject research'' and
``biological studies'' tend to be more collaborative than the topic of
``strategies'', and authors in these are more likely to have at least one
collaborator compared to the topic of ``modeling problems as a PD''.

Topic ``Evolutionary dynamics on networks'' also appears to be a collaborative
topic. It is the topic with smallest number of isolated authors, and has an
average degree of 3.4. In fact the network of the topic is a
sub graph of \(\bar{G}\), the main cluster of $G$. This is discussed in the next
part of this analysis.

\begin{table}[H]
    \centering
    \resizebox{\textwidth}{!}{
\begin{tabular}{lrrrrrrrrrr}
        \toprule
        {} &  \# Nodes &  \# Edges &  \# Isolated nodes &  \% Isolated nodes &  \# Connected components &  Size of largest component &  Av. degree &  \# Communities &  Modularity &  Clustering coeff \\
        \midrule
        Topic A &     1193 &     2137 &                84 &               7.0 &                     333 &                         56 &       3.583 &            334 &       0.983 &             0.715 \\
        Topic B &      727 &     1382 &                45 &               6.2 &                     189 &                         80 &       3.802 &            190 &       0.950 &             0.739 \\
        Topic C &      931 &     1141 &                72 &               7.7 &                     312 &                         29 &       2.451 &            312 &       0.981 &             0.615 \\
        Topic D &      891 &     1509 &                28 &               3.1 &                     185 &                        312 &       3.387 &            193 &       0.917 &             0.692 \\
        Topic E &     1152 &     1964 &               166 &              14.4 &                     461 &                         31 &       3.410 &            461 &       0.926 &             0.602 \\
        \bottomrule
\end{tabular}
    }
    \caption{Network metrics for topic networks.}\label{table:topics_networks}
\end{table}

There are two centrality measures reported in this work, closeness and
betweenness centrality. Closeness centrality is a measure of how easy it is for
an author to reach others, and betweenness centrality is a
measure of how many paths pass through a specific node. All centrality measures 
have values ranging from 0 to 1.

For \(G\) and \(\bar{G}\) the most central authors based on closeness and
betweenness centralities are given by Table~\ref{table:central_authors}. The
most central authors in \(G\) and \(\bar{G}\) are the same. This implies that
the results on centrality heavily rely on the main cluster (as expected). Matjaz Perc is an
author with 83 publications in the data set and the most central authors based
on both centrality measures. The most central authors are fairly similar between
the two measures. The author that appear to be central based on one measure and
not the other are Martin Nowak, Franz Weissing, Jianye Hao, Angel Sanchez and
Valerio Capraro which are central based on betweeness centrality, and 
the opposite is true for Attila Szolnoki, Luo-Luo Jiang Sandro Meloni, Cheng-Yi
Xia and Xiaojie Chen.

\begin{table}[!hbtp]
    \begin{center}
    \resizebox{.9\textwidth}{!}{
\begin{tabular}{llrlrlrlr}
\toprule
& \multicolumn{4}{c}{$G$} & \multicolumn{4}{c}{$\bar{G}$} \\
\midrule
{} &             Name &  Betweenness &             Name &  Closeness &             Name &  Betweenness &             Name &  Closeness \\
\midrule
1  &      Matjaz Perc &        0.013 &      Matjaz Perc &      0.062 &      Matjaz Perc &        0.373 &      Matjaz Perc &      0.330 \\
2  &        Zhen Wang &        0.010 &        Long Wang &      0.057 &        Zhen Wang &        0.279 &        Long Wang &      0.301 \\
3  &        Long Wang &        0.006 &     Yamir Moreno &      0.056 &        Long Wang &        0.170 &     Yamir Moreno &      0.299 \\
4  &     Martin Nowak &        0.006 &  Attila Szolnoki &      0.056 &     Martin Nowak &        0.159 &  Attila Szolnoki &      0.297 \\
5  &    Angel Sanchez &        0.004 &        Zhen Wang &      0.056 &    Angel Sanchez &        0.114 &        Zhen Wang &      0.296 \\
6  &     Yamir Moreno &        0.004 &    Arne Traulsen &      0.053 &     Yamir Moreno &        0.110 &    Arne Traulsen &      0.281 \\
7  &    Arne Traulsen &        0.004 &    Luo-Luo Jiang &      0.053 &    Arne Traulsen &        0.107 &    Luo-Luo Jiang &      0.280 \\
8  &   Franz Weissing &        0.004 &    Sandro Meloni &      0.052 &   Franz Weissing &        0.101 &    Sandro Meloni &      0.278 \\
9  &       Jianye Hao &        0.003 &     Cheng-Yi Xia &      0.052 &       Jianye Hao &        0.094 &     Cheng-Yi Xia &      0.276 \\
10 &  Valerio Capraro &        0.003 &     Xiaojie Chen &      0.052 &  Valerio Capraro &        0.093 &     Xiaojie Chen &      0.276 \\
\bottomrule
\end{tabular}
    }
\end{center}
\caption{10 most central authors based on betweenness and closeness centralities
for \(G\) and \(\bar{G}\).}\label{table:central_authors}
\end{table}

The
centrality measures for the topic networks have also been estimated and are
given in Tables~\ref{table:central_authors_topics_bt}
and~\ref{table:central_authors_topics_cl}. The centrality measure for the
topics' networks are low except from the case of Topic D. From the
list of names it is obvious that the most central authors of Topic D are part of \(\bar{G}\), and that
the network of evolutionary dynamics on networks is a sub network of \(\bar{G}\).

This confirms the result that the most central authors of 
the co-authorship network are the authors of the main cluster of \(G\).
The fact that most authors of the main cluster are primarily publishing in
evolutionary dynamics on networks indicates that publishing in this specific
topic differs from the other topics covered in this manuscript. It could also
indicate that authors publishing in evolutionary dynamics are more similar to other
disciplines as they can collaborate with them more.

\newcolumntype{g}{>{\columncolor{Gray}}l}
\begin{table}[!hbtp]
    \begin{center}
    \resizebox{.9\textwidth}{!}{
\begin{tabular}{lggllggllgg}
\toprule
& \multicolumn{2}{g}{Topic A} & \multicolumn{2}{c}{Topic B} & \multicolumn{2}{g}{Topic C} & \multicolumn{2}{c}{Topic D} & \\
\midrule
{} &                 Name &  Betweeness &             Name &  Betweeness &             Name &  Betweeness &              Name &  Betweeness &                  Name &  Betweeness \\
\midrule
1  &           David Rand &       0.001 &        Long Wang &       0.006 &   Daniel Ashlock &       0.001 &       Matjaz Perc &       0.062 &             Zengru Di &         0.0 \\
2  &      Valerio Capraro &       0.001 &    Luo-Luo Jiang &       0.004 &      Matjaz Perc &       0.000 &     Luo-Luo Jiang &       0.036 &             Jian Yang &         0.0 \\
3  &        Angel Sanchez &       0.000 &     Martin Nowak &       0.004 &       Karl Tuyls &       0.000 &      Yamir Moreno &       0.031 &  Yevgeniy Vorobeychik &         0.0 \\
4  &              Feng Fu &       0.000 &      Matjaz Perc &       0.003 &  Philip Hingston &       0.000 &  Christoph Hauert &       0.027 &       Otavio Teixeira &         0.0 \\
5  &         Martin Nowak &       0.000 &  Attila Szolnoki &       0.002 &     Eun-Youn Kim &       0.000 &         Long Wang &       0.023 &      Roberto Oliveira &         0.0 \\
6  &  Nicholas Christakis &       0.000 &  Christian Hilbe &       0.002 &    Wendy Ashlock &       0.000 &         Zhen Wang &       0.023 &              M. Nowak &         0.0 \\
7  &   Pablo Branas-Garza &       0.000 &     Yamir Moreno &       0.002 &  Attila Szolnoki &       0.000 &      Han-Xin Yang &       0.022 &             M. Harper &         0.0 \\
8  &     Toshio Yamagishi &       0.000 &     Xiaojie Chen &       0.002 &       Seung Baek &       0.000 &      Martin Nowak &       0.019 &              Xiao Han &         0.0 \\
9  &         James Fowler &       0.000 &    Arne Traulsen &       0.002 &     Martin Nowak &       0.000 &     Angel Sanchez &       0.016 &            Zhesi Shen &         0.0 \\
10 &            Long Wang &       0.000 &        Zhen Wang &       0.002 &    Thore Graepel &       0.000 &       Zhihai Rong &       0.015 &           Wen-Xu Wang &         0.0 \\
\bottomrule
\end{tabular}
}
\end{center}
\caption{10 most central authors based on betweeness centrality
for topics' networks.}\label{table:central_authors_topics_bt}
\end{table}

\newcolumntype{g}{>{\columncolor{Gray}}l}
\begin{table}[!hbtp]
    \begin{center}
\resizebox{.9\textwidth}{!}{
\begin{tabular}{lggllggllgg}
    \toprule
    & \multicolumn{2}{g}{Topic A} & \multicolumn{2}{c}{Topic B} & \multicolumn{2}{g}{Topic C} & \multicolumn{2}{c}{Topic D} & \\
    \midrule
    {} &                 Name &  Closeness &               Name &  Closeness &                 Name &  Closeness &             Name &  Closeness &             Name &  Closeness \\
    \midrule
    1  &           David Rand &      0.026 &          Long Wang &      0.042 &           Karl Tuyls &      0.021 &      Matjaz Perc &      0.122 &  Stefanie Widder &      0.026 \\
    2  &      Valerio Capraro &      0.022 &        Matjaz Perc &      0.039 &        Thore Graepel &      0.019 &        Zhen Wang &      0.107 &   Rosalind Allen &      0.026 \\
    3  &       Jillian Jordan &      0.021 &    Attila Szolnoki &      0.039 &           Joel Leibo &      0.018 &        Long Wang &      0.105 &  Thomas Pfeiffer &      0.026 \\
    4  &  Nicholas Christakis &      0.020 &       Martin Nowak &      0.038 &        Edward Hughes &      0.017 &     Yamir Moreno &      0.103 &    Thomas Curtis &      0.026 \\
    5  &         James Fowler &      0.019 &  Olivier Tenaillon &      0.037 &     Matthew Phillips &      0.017 &    Luo-Luo Jiang &      0.102 &     Carsten Wiuf &      0.026 \\
    6  &         Martin Nowak &      0.019 &       Xiaojie Chen &      0.036 &  Edgar Duenez-Guzman &      0.017 &  Attila Szolnoki &      0.102 &    William Sloan &      0.026 \\
    7  &        Angel Sanchez &      0.018 &             Bin Wu &      0.036 &    Antonio Castaneda &      0.017 &     Gyorgy Szabo &      0.101 &     Otto Cordero &      0.026 \\
    8  &      Samuel Arbesman &      0.018 &      Yanling Zhang &      0.035 &         Iain Dunning &      0.017 &     Xiaojie Chen &      0.100 &        Sam Brown &      0.026 \\
    9  &    Gordon Kraft-Todd &      0.018 &            Feng Fu &      0.035 &             Tina Zhu &      0.017 &    Guangming Xie &      0.100 &     Babak Momeni &      0.026 \\
    10 &        Akihiro Nishi &      0.018 &         David Rand &      0.035 &          Kevin Mckee &      0.017 &     Lucas Wardil &      0.100 &     Wenying Shou &      0.026 \\
    \bottomrule
\end{tabular}
}
\end{center}
\caption{10 most central authors based on closeness centrality
for topics' networks.}\label{table:central_authors_topics_cl}
\end{table}

The distributions of both centrality measures for all the networks of this
work are given in the Appendix~\ref{appendix:distributions}.

\section{Conclusion}\label{section:conclusion}

This manuscript has explored the research topics in the publications of the
Iterated Prisoner's Dilemma, and moreover, the authors' collaborative behaviour
and their centrality. This was achieved by
applying network theoretic approaches and a LDA algorithm to a total of 2422
publications.
The data collection and an initial analysis of the data set were covered in
Section~\ref{section:methodology}. The analysis  demonstrated that the PD is a
field that continues to attract academic attention and publications.

In Section~\ref{subsection:research_topics} LDA was applied to the data set to
identify topics on which researchers have been publishing. The five topics in
the PD publications identified by the data set of this work are human subject
research, biological studies, strategies, evolutionary dynamics on networks and
modeling problems as a PD. These 5 topics nicely summarise PD research. They
highlight the interdisciplinarity of the field; how it brings together applied
modeling of real world situations (biological studies and modeling problems as a
PD) and more theoretical notions such as evolutionary dynamics and optimality of
strategies. A temporal analysis explored how relevant these topics have been
over the course of time, and it revealed that even though they were not
necessarily always the most discussed topics there were still being explored
by researchers.

The collaborative behaviour of the field was explored in
Section~\ref{subsection:co_authorship} investigated the co-authorship
network. It was concluded that the field is a collaborative field, where authors
are likely to write with a collaborator's co-authors and on average an author
has 4 co-authors. The results were compared to the networks of two other game
theoretic fields, and it was shown that the PD network is relatively more collaborative. The
authors however, tend to collaborate with authors from one community, but not
many authors are involved in multiple communities. This might be an effect of
academic research, and it might not be true just for the field of the PD.

Exploring the centrality of authors showed that the most central author of this
manuscript is Matjaz Perc. More importantly, it was shown that most central
authors of the network were the authors connected to the main cluster.
Interestingly, it was uncovered that these authors were the most central due to
their publication on a single topic alone. That was the topic of “evolutionary
dynamics on networks”. There appears to be more collaboration and more influence
in the publications on evolutionary dynamics. The authors are most likely to
gain from their position, and come across as the more important authors in the
field. Though it is not clear as to why, attention should be paid to the
collaborative behaviour of authors of “evolutionary dynamics on networks”.

The study of the PD is the study of cooperation and investigating the
cooperative behaviours of authors is what this work has aimed to achieve.
Interesting areas of future work would include extending this analysis to more
game theoretic sub fields, to evaluate whether the results remain the same.
Moreover, the networks of this work were created by not taking into account the
strength of ties. The strength of ties could be analysed to map multiple
collaborations between two nodes. However, a preliminary assessment showed that
the presented results do not change.

Both the software~\citep{nikoleta_2017} and the data~\citep{pd_data_2018} used in
the manuscript have been archived and are available to be used by other
researchers.

\section*{Competing Interests}

The authors declare no competing interests.

\section*{Acknowledgements}

The authors would like to thank the anonymous reviewers for their comments which
helped improve the paper.

\section*{Data Availability}

The datasets generated and analysed during the current study are available
in~\citep{anarchy_data_2018, auction_data_2018, pd_data_2018}.

\section*{Figures Legends}

\begin{itemize}
  \item Figure~\ref{figure:colection_process}. The generating process of the
  data set~\citep{pd_data_2018}.
  \item Figure~\ref{fig:timeseries}. Number of articles published on the PD
  1951-2018 (on a log scale), with a fitted exponential line, and aforecast for
  2017-2022.
  \item Figure~\ref{figure:models_scores}. Coherence and exclusivity for LDA
  models over the number of topics.
  \item Figure~\ref{fig:number_of_articles_per_topic}. Number of articles per
  topic over the years (on a logged scale).
  \item Figure~\ref{fig:median_percentage_contribution_over_time}. Maximum
  percentage contributions (\(c^*\)) over the time periods,
  for the LDA models for the entire data set for \(n\) equal to 5, 6
  and for the models of Table~\ref{table:topics_per_year}.
\end{itemize}

\appendix

\section{Topic modeling results over time periods}\label{appendix:topics_per_year}

\begin{table}[!hbtp]
    \begin{center}
    \resizebox{\textwidth}{!}{
        \begin{tabular}{llccc}
            \toprule
                Period &  Topic &                                                                                                Topic Keywords & Num of Documents & Percentage of Documents \\
            \midrule
             1951-1965 &               1 &                 problem, technology, divert, euler, subsystem, requirement, trace, technique, system, untried &                3 &                   0.375 \\
             1951-1965 &               2 &            interpret, requirement, programme, evolution, article, increase, policy, system, trace, technology &                2 &                    0.25 \\
             1951-1965 &               3 &          equipment, agency, conjecture, development, untried, programme, trend, technology, weapon, technique &                1 &                   0.125 \\
             1951-1965 &               4 &                 variation, celebrated, trend, untried, change, involve, month, technique, subsystem, research &                1 &                   0.125 \\
             1951-1965 &               5 &                           give, good, modern, trace, technique, ambiguity, problem, trend, technology, system &                1 &                   0.125 \\
             \midrule
             1951-1973 &               1 &                           study, shock, cooperative, money, part, vary, investigate, good, receive, equipment &               12 &                  0.3243 \\
             1951-1973 &               2 &          cooperation, level, significantly, sequence, reward, provoke, descriptive, principal, display, argue &                4 &                  0.1081 \\
             1951-1973 &               3 &               player, make, effect, triad, experimental, motivation, dominate, hypothesis, instruction, trend &                3 &                  0.0811 \\
             1951-1973 &               4 &                                           ss, sex, male, female, dyad, design, suggest, college, factor, tend &                3 &                  0.0811 \\
             1951-1973 &               5 &               result, research, format, change, operational, analysis, relate, understanding, decision, money &                2 &                  0.0541 \\
             1951-1973 &               6 &                          condition, give, high, treatment, conflict, cc, real, original, replication, promote &                2 &                  0.0541 \\
             1951-1973 &               7 &              group, competitive, show, interpret, scale, compete, escalation, free, variable, individualistic &                2 &                  0.0541 \\
             1951-1973 &               8 &                        outcome, strategy, choice, type, pdg, difference, dummy, conclude, compare, consistent &                2 &                  0.0541 \\
             1951-1973 &               9 &                   game, difference, pair, approach, behavior, person, weapon, occur, advantaged, differential &                2 &                  0.0541 \\
             1951-1973 &              10 &                    response, present, dilemma, influence, cooperate, bias, point, amount, participate, factor &                2 &                  0.0541 \\
             1951-1973 &              11 &                       trial, problem, previous, involve, prisoner, experiment, follow, tit, increase, initial &                1 &                   0.027 \\
             1951-1973 &              12 &                           matrix, behavior, rational, black, model, research, broad, distance, complex, trace &                1 &                   0.027 \\
             1951-1973 &              13 &                    play, finding, individual, noncooperative, white, nature, race, ratio, represent, prisoner &                1 &                   0.027 \\
             \midrule
             1951-1980 &               1 &                                      play, trial, group, follow, white, interpret, scale, black, trend, small &               14 &                    0.25 \\
             1951-1980 &               2 &                              outcome, level, effect, type, dyad, vary, pdg, participate, understanding, arise &                9 &                  0.1607 \\
             1951-1980 &               3 &         game, strategy, cooperation, significant, difference, sentence, text, occur, differential, hypothesis &                4 &                  0.0714 \\
             1951-1980 &               4 &                        male, female, find, result, sex, subject, experimental, situation, treatment, computer &                4 &                  0.0714 \\
             1951-1980 &               5 &                         research, problem, influence, matrix, format, model, analysis, year, crime, equipment &                4 &                  0.0714 \\
             1951-1980 &               6 &                                    condition, dilemma, bias, free, attempt, book, year, dummy, prison, design &                4 &                  0.0714 \\
             1951-1980 &               7 &                    variable, result, factor, individual, ability, triad, half, migration, change, investigate &                3 &                  0.0536 \\
             1951-1980 &               8 &                 show, present, suggest, rational, compete, approach, characteristic, examine, person, conduct &                3 &                  0.0536 \\
             1951-1980 &               9 &                         behavior, high, finding, relate, obtain, assistance, ratio, good, weapon, competition &                3 &                  0.0536 \\
             1951-1980 &              10 &                               ss, shock, money, competitive, part, difference, pair, amount, man, information &                3 &                  0.0536 \\
             1951-1980 &              11 &             player, conflict, theory, decision, determine, produce, maker, cooperate, specialist, programming &                2 &                  0.0357 \\
             1951-1980 &              12 &            study, prisoner, make, response, experiment, noncooperative, standard, separate, conclude, initial &                2 &                  0.0357 \\
             1951-1980 &              13 &                       give, cooperative, choice, cognitive, real, operational, set, subject, ascribe, concern &                1 &                  0.0179 \\
             \midrule
             1951-1988 &               1 &                     trial, difference, find, choice, significant, competitive, effect, triad, interact, occur &               24 &                  0.2553 \\
             1951-1988 &               2 &                                            ss, shock, money, pair, response, part, high, tit, receive, amount &               13 &                  0.1383 \\
             1951-1988 &               3 &                         suggest, paper, case, debate, view, achieve, framework, natural, assumption, finitely &               10 &                  0.1064 \\
             1951-1988 &               4 &                     prisoner, dilemma, behavior, model, present, involve, person, increase, trust, experiment &                8 &                  0.0851 \\
             1951-1988 &               5 &                                   game, player, show, approach, repeat, previous, move, tat, related, include &                8 &                  0.0851 \\
             1951-1988 &               6 &                cooperation, level, mutual, equilibrium, standard, provide, information, human, real, question &                6 &                  0.0638 \\
             1951-1988 &               7 &                      play, result, male, subject, female, cooperative, sex, experimental, treatment, computer &                5 &                  0.0532 \\
             1951-1988 &               8 &                        research, study, variable, ability, factor, conflict, matrix, year, student, interpret &                4 &                  0.0426 \\
             1951-1988 &               9 &                                         problem, group, small, scale, social, issue, large, base, bias, party &                4 &                  0.0426 \\
             1951-1988 &              10 &                          game, strategy, outcome, type, cooperate, ethical, pdg, explain, dependent, separate &                4 &                  0.0426 \\
             1951-1988 &              11 &              give, condition, individual, major, dyad, behaviour, produce, conflict, assistance, collectively &                3 &                  0.0319 \\
             1951-1988 &              12 &                        situation, iterate, statement, rational, card, side, paradox, true, consequence, front &                2 &                  0.0213 \\
             1951-1988 &              13 &                               inflation, hypothesis, rate, run, change, demand, nominal, cost, output, growth &                2 &                  0.0213 \\
             1951-1988 &              14 &                                     theory, make, analysis, decision, system, examine, work, soft, lead, hard &                1 &                  0.0106 \\
             \midrule
             1951-1995 &               1 &                            strategy, population, evolution, iterate, tit, opponent, evolve, dynamic, set, tat &               31 &                  0.1732 \\
             1951-1995 &               2 &                 game, repeat, assumption, rule, person, equilibrium, general, finitely, indefinitely, analyze &               24 &                  0.1341 \\
             1951-1995 &               3 &                            inflation, long, rate, hypothesis, run, policy, cost, nominal, demand, programming &               20 &                  0.1117 \\
             1951-1995 &               4 &            condition, outcome, trial, find, difference, cooperation, experiment, level, significant, response &               15 &                  0.0838 \\
             1951-1995 &               5 &                     rational, result, receive, statement, money, paradox, shock, iterate, consequence, common &               14 &                  0.0782 \\
             1951-1995 &               6 &             cooperation, show, competitive, high, probability, conflict, simulation, altruism, yield, natural &               14 &                  0.0782 \\
             1951-1995 &               7 &                           prisoner, dilemma, give, point, defect, form, cooperator, increase, relate, ethical &               10 &                  0.0559 \\
             1951-1995 &               8 &                       player, give, decision, provide, cooperative, game, previous, pair, determine, interact &                9 &                  0.0503 \\
             1951-1995 &               9 &                          play, cooperate, result, male, subject, female, time, relationship, suggest, student &                8 &                  0.0447 \\
             1951-1995 &              10 &                                   problem, group, theory, good, approach, society, large, scale, issue, level &                8 &                  0.0447 \\
             1951-1995 &              11 &            study, situation, behaviour, computer, argue, change, implication, characteristic, real, associate &                8 &                  0.0447 \\
             1951-1995 &              12 &                        model, paper, behavior, examine, present, mutual, expectation, develop, type, variable &                7 &                  0.0391 \\
             1951-1995 &              13 &                                   make, research, system, analysis, choice, work, base, relation, world, wide &                6 &                  0.0335 \\
             1951-1995 &              14 &               individual, social, behavior, standard, choose, evolutionary, partner, payoff, defection, small &                5 &                  0.0279 \\
             \midrule
             1951-2003 &               1 &                                    game, player, dilemma, prisoner, theory, give, paper, make, group, problem &              151 &                  0.4266 \\
             1951-2003 &               2 &                         cooperation, result, play, show, cooperate, condition, cooperative, high, level, time &              106 &                  0.2994 \\
             1951-2003 &               3 &                  strategy, model, agent, study, behavior, individual, population, evolutionary, state, player &               97 &                   0.274 \\
             \midrule
             1951-2010 &               1 &                                  model, theory, paper, base, make, present, problem, provide, human, decision &              325 &                  0.3454 \\
             1951-2010 &               2 &                                   game, strategy, player, agent, play, dilemma, system, behavior, show, state &              322 &                  0.3422 \\
             1951-2010 &               3 &  cooperation, network, study, population, individual, evolutionary, social, evolution, interaction, structure &              294 &                  0.3124 \\
             \midrule
             1951-2018 &               1 &                              model, theory, system, base, paper, problem, propose, present, approach, provide &              556 &                  0.2251 \\
             1951-2018 &               2 &                        behavior, social, human, decision, study, experiment, make, suggest, result, behaviour &              482 &                  0.1951 \\
             1951-2018 &               3 &                     individual, group, good, social, punishment, level, cost, mechanism, dilemma, cooperative &              428 &                  0.1733 \\
             1951-2018 &               4 &                            game, strategy, player, agent, play, dilemma, state, prisoner, payoff, equilibrium &              380 &                  0.1538 \\
             1951-2018 &               5 &                 population, evolutionary, dynamic, model, selection, result, evolution, evolve, show, process &              351 &                  0.1421 \\
             1951-2018 &               6 &       cooperation, network, interaction, structure, study, evolution, find, behavior, cooperative, simulation &              273 &                  0.1105 \\
            \bottomrule
            \end{tabular}            
    }
    \end{center}
    \caption{Topic modeling results for the cumulative data sets over the
    periods: 1951-1965, 1951-1973, 1951-1980, 1951-1988, 1951-1995, 1951-2003, 1951-2010,
    1951-2018. The number of topics \(n\) for each period is given in the column
    ``Topic". For example in the period 1951-1980 the selected \(n\) was 13. The
    number of topics here were chosen only based on the coherence score.
    The number of documents and the percentage of documents assigned to each
    topic, for each period is also given.}\label{table:topics_per_year}
\end{table}

\section{Cumulative Networks Metrics}\label{appendix:tables}

\subsection{Collaborativeness metrics for cumulative graphs, \(\tilde{G} \subseteq G\)}
\begin{table}[!hbtp]
    \centering
    \resizebox{.8\textwidth}{!}{
        \begin{tabular}{lrrrrrrrrrr}
            \toprule
            {} &  \# Nodes &  \# Edges &  \# Isolated nodes &  \% Isolated nodes &  \# Connected components &  Size of largest component &  Av. degree &  \# Communities &  Modularity &  Clustering coeff \\
            \midrule
            Period 0  &       11 &        3 &                 5 &              45.5 &                       8 &                          2 &       0.545 &              8 &       0.667 &             0.000 \\
            Period 1  &       14 &        4 &                 6 &              42.9 &                      10 &                          2 &       0.571 &             10 &       0.750 &             0.000 \\
            Period 2  &       27 &       15 &                 8 &              29.6 &                      16 &                          5 &       1.111 &             16 &       0.684 &             0.160 \\
            Period 3  &       29 &       17 &                 9 &              31.0 &                      17 &                          6 &       1.172 &             17 &       0.630 &             0.172 \\
            Period 4  &       32 &       18 &                10 &              31.2 &                      19 &                          6 &       1.125 &             19 &       0.667 &             0.156 \\
            Period 5  &       43 &       28 &                10 &              23.3 &                      23 &                          6 &       1.302 &             23 &       0.827 &             0.326 \\
            Period 6  &       49 &       34 &                10 &              20.4 &                      25 &                          6 &       1.388 &             25 &       0.867 &             0.408 \\
            Period 7  &       52 &       35 &                11 &              21.2 &                      27 &                          6 &       1.346 &             27 &       0.873 &             0.385 \\
            Period 8  &       54 &       35 &                13 &              24.1 &                      29 &                          6 &       1.296 &             29 &       0.873 &             0.370 \\
            Period 9  &       54 &       35 &                13 &              24.1 &                      29 &                          6 &       1.296 &             29 &       0.873 &             0.370 \\
            Period 10 &       59 &       36 &                16 &              27.1 &                      33 &                          6 &       1.220 &             33 &       0.880 &             0.339 \\
            Period 11 &       60 &       36 &                17 &              28.3 &                      34 &                          6 &       1.200 &             34 &       0.880 &             0.333 \\
            Period 12 &       63 &       40 &                17 &              27.0 &                      34 &                          6 &       1.270 &             34 &       0.884 &             0.339 \\
            Period 13 &       65 &       40 &                19 &              29.2 &                      36 &                          6 &       1.231 &             36 &       0.884 &             0.328 \\
            Period 14 &       69 &       46 &                20 &              29.0 &                      37 &                          6 &       1.333 &             37 &       0.889 &             0.360 \\
            Period 15 &       71 &       46 &                22 &              31.0 &                      39 &                          6 &       1.296 &             39 &       0.889 &             0.350 \\
            Period 16 &       75 &       47 &                24 &              32.0 &                      42 &                          6 &       1.253 &             42 &       0.894 &             0.331 \\
            Period 17 &       80 &       47 &                29 &              36.2 &                      47 &                          6 &       1.175 &             47 &       0.894 &             0.310 \\
            Period 18 &       84 &       47 &                33 &              39.3 &                      51 &                          6 &       1.119 &             51 &       0.894 &             0.296 \\
            Period 19 &       92 &       48 &                39 &              42.4 &                      58 &                          6 &       1.043 &             58 &       0.898 &             0.270 \\
            Period 20 &      101 &       52 &                43 &              42.6 &                      64 &                          6 &       1.030 &             64 &       0.909 &             0.276 \\
            Period 21 &      114 &       62 &                44 &              38.6 &                      70 &                          6 &       1.088 &             70 &       0.926 &             0.279 \\
            Period 22 &      119 &       64 &                45 &              37.8 &                      73 &                          6 &       1.076 &             73 &       0.930 &             0.268 \\
            Period 23 &      129 &       69 &                48 &              37.2 &                      79 &                          6 &       1.070 &             79 &       0.937 &             0.270 \\
            Period 24 &      140 &       72 &                55 &              39.3 &                      87 &                          6 &       1.029 &             87 &       0.941 &             0.249 \\
            Period 25 &      154 &       81 &                60 &              39.0 &                      95 &                          6 &       1.052 &             95 &       0.947 &             0.252 \\
            Period 26 &      179 &       95 &                71 &              39.7 &                     111 &                          6 &       1.061 &            111 &       0.955 &             0.273 \\
            Period 27 &      192 &      102 &                74 &              38.5 &                     118 &                          6 &       1.062 &            118 &       0.960 &             0.270 \\
            Period 28 &      199 &      105 &                75 &              37.7 &                     122 &                          6 &       1.055 &            122 &       0.962 &             0.260 \\
            Period 29 &      214 &      115 &                79 &              36.9 &                     130 &                          6 &       1.075 &            130 &       0.966 &             0.284 \\
            Period 30 &      255 &      140 &                85 &              33.3 &                     151 &                          6 &       1.098 &            151 &       0.973 &             0.275 \\
            Period 31 &      288 &      169 &                92 &              31.9 &                     166 &                          6 &       1.174 &            166 &       0.977 &             0.304 \\
            Period 32 &      319 &      195 &                96 &              30.1 &                     179 &                          6 &       1.223 &            179 &       0.979 &             0.327 \\
            Period 33 &      360 &      235 &               103 &              28.6 &                     198 &                          7 &       1.306 &            198 &       0.977 &             0.334 \\
            Period 34 &      411 &      278 &               112 &              27.3 &                     222 &                          7 &       1.353 &            222 &       0.979 &             0.349 \\
            Period 35 &      459 &      310 &               118 &              25.7 &                     246 &                          7 &       1.351 &            246 &       0.982 &             0.343 \\
            Period 36 &      521 &      370 &               124 &              23.8 &                     269 &                         10 &       1.420 &            269 &       0.983 &             0.355 \\
            Period 37 &      621 &      476 &               130 &              20.9 &                     303 &                         19 &       1.533 &            303 &       0.985 &             0.393 \\
            Period 38 &      738 &      603 &               141 &              19.1 &                     344 &                         22 &       1.634 &            344 &       0.987 &             0.422 \\
            Period 39 &      904 &      877 &               157 &              17.4 &                     394 &                         25 &       1.940 &            394 &       0.985 &             0.467 \\
            Period 40 &     1062 &     1170 &               164 &              15.4 &                     432 &                         33 &       2.203 &            433 &       0.985 &             0.498 \\
            Period 41 &     1232 &     1442 &               178 &              14.4 &                     480 &                         71 &       2.341 &            482 &       0.982 &             0.515 \\
            Period 42 &     1429 &     1936 &               195 &              13.6 &                     531 &                        133 &       2.710 &            535 &       0.965 &             0.538 \\
            Period 43 &     1695 &     2375 &               214 &              12.6 &                     607 &                        157 &       2.802 &            610 &       0.970 &             0.564 \\
            Period 44 &     1980 &     2865 &               223 &              11.3 &                     677 &                        209 &       2.894 &            680 &       0.968 &             0.589 \\
            Period 45 &     2300 &     3420 &               244 &              10.6 &                     754 &                        322 &       2.974 &            760 &       0.965 &             0.602 \\
            Period 46 &     2643 &     3971 &               265 &              10.0 &                     845 &                        399 &       3.005 &            856 &       0.962 &             0.618 \\
            Period 47 &     3106 &     4877 &               278 &               9.0 &                     933 &                        504 &       3.140 &            947 &       0.965 &             0.639 \\
            Period 48 &     3664 &     6532 &               309 &               8.4 &                    1045 &                        613 &       3.566 &           1058 &       0.964 &             0.659 \\
            Period 49 &     3938 &     7072 &               322 &               8.2 &                    1098 &                        706 &       3.592 &           1115 &       0.965 &             0.664 \\
            Period 50 &     4221 &     7642 &               338 &               8.0 &                    1157 &                        796 &       3.621 &           1177 &       0.965 &             0.666 \\
            \bottomrule
            \end{tabular}
}
    \caption{Collaborativeness metrics for cumulative graphs, \(\tilde{G} \subseteq G\).}
\end{table}

\newpage

\subsection{Collaborativeness metrics for cumulative graphs' main clusters, \(\tilde{G} \subseteq \bar{G}\)}
\begin{table}[!hbtp]
    \centering
    \resizebox{.8\textwidth}{!}{
        \begin{tabular}{lrrrrrrrrrr}
            \toprule
            {} &  \# Nodes &  \# Edges &  \# Isolated nodes &  \% Isolated nodes &  \# Connected components &  Size of largest component &  Av. degree &  \# Communities &  Modularity &  Clustering coeff \\
            \midrule
            0  &        2 &        1 &                 0 &               0.0 &                       1 &                          2 &       1.000 &              1 &       0.000 &             0.000 \\
            1  &        2 &        1 &                 0 &               0.0 &                       1 &                          2 &       1.000 &              1 &       0.000 &             0.000 \\
            2  &        5 &        8 &                 0 &               0.0 &                       1 &                          5 &       3.200 &              1 &       0.000 &             0.867 \\
            3  &        6 &       10 &                 0 &               0.0 &                       1 &                          6 &       3.333 &              2 &       0.020 &             0.833 \\
            4  &        6 &       10 &                 0 &               0.0 &                       1 &                          6 &       3.333 &              2 &       0.020 &             0.833 \\
            5  &        6 &       10 &                 0 &               0.0 &                       1 &                          6 &       3.333 &              2 &       0.020 &             0.833 \\
            6  &        6 &       10 &                 0 &               0.0 &                       1 &                          6 &       3.333 &              2 &       0.020 &             0.833 \\
            7  &        6 &       10 &                 0 &               0.0 &                       1 &                          6 &       3.333 &              2 &       0.020 &             0.833 \\
            8  &        6 &       10 &                 0 &               0.0 &                       1 &                          6 &       3.333 &              2 &       0.020 &             0.833 \\
            9  &        6 &       10 &                 0 &               0.0 &                       1 &                          6 &       3.333 &              2 &       0.020 &             0.833 \\
            10 &        6 &       10 &                 0 &               0.0 &                       1 &                          6 &       3.333 &              2 &       0.020 &             0.833 \\
            11 &        6 &       10 &                 0 &               0.0 &                       1 &                          6 &       3.333 &              2 &       0.020 &             0.833 \\
            12 &        6 &       10 &                 0 &               0.0 &                       1 &                          6 &       3.333 &              2 &       0.020 &             0.833 \\
            13 &        6 &       10 &                 0 &               0.0 &                       1 &                          6 &       3.333 &              2 &       0.020 &             0.833 \\
            14 &        6 &       10 &                 0 &               0.0 &                       1 &                          6 &       3.333 &              2 &       0.020 &             0.833 \\
            15 &        6 &       10 &                 0 &               0.0 &                       1 &                          6 &       3.333 &              2 &       0.020 &             0.833 \\
            16 &        6 &       10 &                 0 &               0.0 &                       1 &                          6 &       3.333 &              2 &       0.020 &             0.833 \\
            17 &        6 &       10 &                 0 &               0.0 &                       1 &                          6 &       3.333 &              2 &       0.020 &             0.833 \\
            18 &        6 &       10 &                 0 &               0.0 &                       1 &                          6 &       3.333 &              2 &       0.020 &             0.833 \\
            19 &        6 &       10 &                 0 &               0.0 &                       1 &                          6 &       3.333 &              2 &       0.020 &             0.833 \\
            20 &        6 &       10 &                 0 &               0.0 &                       1 &                          6 &       3.333 &              2 &       0.020 &             0.833 \\
            21 &        6 &       10 &                 0 &               0.0 &                       1 &                          6 &       3.333 &              2 &       0.020 &             0.833 \\
            22 &        6 &       10 &                 0 &               0.0 &                       1 &                          6 &       3.333 &              2 &       0.020 &             0.833 \\
            23 &        6 &       10 &                 0 &               0.0 &                       1 &                          6 &       3.333 &              2 &       0.020 &             0.833 \\
            24 &        6 &       10 &                 0 &               0.0 &                       1 &                          6 &       3.333 &              2 &       0.020 &             0.833 \\
            25 &        6 &       10 &                 0 &               0.0 &                       1 &                          6 &       3.333 &              2 &       0.020 &             0.833 \\
            26 &        6 &       10 &                 0 &               0.0 &                       1 &                          6 &       3.333 &              2 &       0.020 &             0.833 \\
            27 &        6 &       10 &                 0 &               0.0 &                       1 &                          6 &       3.333 &              2 &       0.020 &             0.833 \\
            28 &        6 &       10 &                 0 &               0.0 &                       1 &                          6 &       3.333 &              2 &       0.020 &             0.833 \\
            29 &        6 &       10 &                 0 &               0.0 &                       1 &                          6 &       3.333 &              2 &       0.020 &             0.833 \\
            30 &        6 &       10 &                 0 &               0.0 &                       1 &                          6 &       3.333 &              2 &       0.020 &             0.833 \\
            31 &        6 &       10 &                 0 &               0.0 &                       1 &                          6 &       3.333 &              2 &       0.020 &             0.833 \\
            32 &        6 &       10 &                 0 &               0.0 &                       1 &                          6 &       3.333 &              2 &       0.020 &             0.833 \\
            33 &        7 &       21 &                 0 &               0.0 &                       1 &                          7 &       6.000 &              1 &       0.000 &             1.000 \\
            34 &        7 &       21 &                 0 &               0.0 &                       1 &                          7 &       6.000 &              1 &       0.000 &             1.000 \\
            35 &        7 &       21 &                 0 &               0.0 &                       1 &                          7 &       6.000 &              1 &       0.000 &             1.000 \\
            36 &       10 &       13 &                 0 &               0.0 &                       1 &                         10 &       2.600 &              2 &       0.376 &             0.553 \\
            37 &       19 &       28 &                 0 &               0.0 &                       1 &                         19 &       2.947 &              3 &       0.544 &             0.730 \\
            38 &       22 &       35 &                 0 &               0.0 &                       1 &                         22 &       3.182 &              4 &       0.541 &             0.720 \\
            39 &       25 &       39 &                 0 &               0.0 &                       1 &                         25 &       3.120 &              5 &       0.558 &             0.686 \\
            40 &       33 &       62 &                 0 &               0.0 &                       1 &                         33 &       3.758 &              4 &       0.623 &             0.736 \\
            41 &       71 &      148 &                 0 &               0.0 &                       1 &                         71 &       4.169 &              6 &       0.726 &             0.698 \\
            42 &      133 &      387 &                 0 &               0.0 &                       1 &                        133 &       5.820 &              7 &       0.726 &             0.749 \\
            43 &      157 &      465 &                 0 &               0.0 &                       1 &                        157 &       5.924 &              8 &       0.721 &             0.725 \\
            44 &      209 &      611 &                 0 &               0.0 &                       1 &                        209 &       5.847 &             13 &       0.738 &             0.737 \\
            45 &      322 &      892 &                 0 &               0.0 &                       1 &                        322 &       5.540 &             16 &       0.780 &             0.743 \\
            46 &      399 &     1109 &                 0 &               0.0 &                       1 &                        399 &       5.559 &             15 &       0.792 &             0.742 \\
            47 &      504 &     1368 &                 0 &               0.0 &                       1 &                        504 &       5.429 &             21 &       0.809 &             0.751 \\
            48 &      613 &     1677 &                 0 &               0.0 &                       1 &                        613 &       5.471 &             24 &       0.820 &             0.761 \\
            49 &      706 &     1935 &                 0 &               0.0 &                       1 &                        706 &       5.482 &             28 &       0.832 &             0.772 \\
            50 &      796 &     2214 &                 0 &               0.0 &                       1 &                        796 &       5.563 &             29 &       0.843 &             0.773 \\
            \bottomrule
            \end{tabular}
            }
    \caption{Collaborativeness metrics for cumulative graphs' main clusters, \(\tilde{G} \subseteq \bar{G}\).}
\end{table}

\section{Centrality Measures Distributions}

\subsection{Distributions for \(G\) and \(\bar{G}\)}

\begin{figure}[!hbtp]
    \centering
    \includegraphics[width=.8\textwidth]{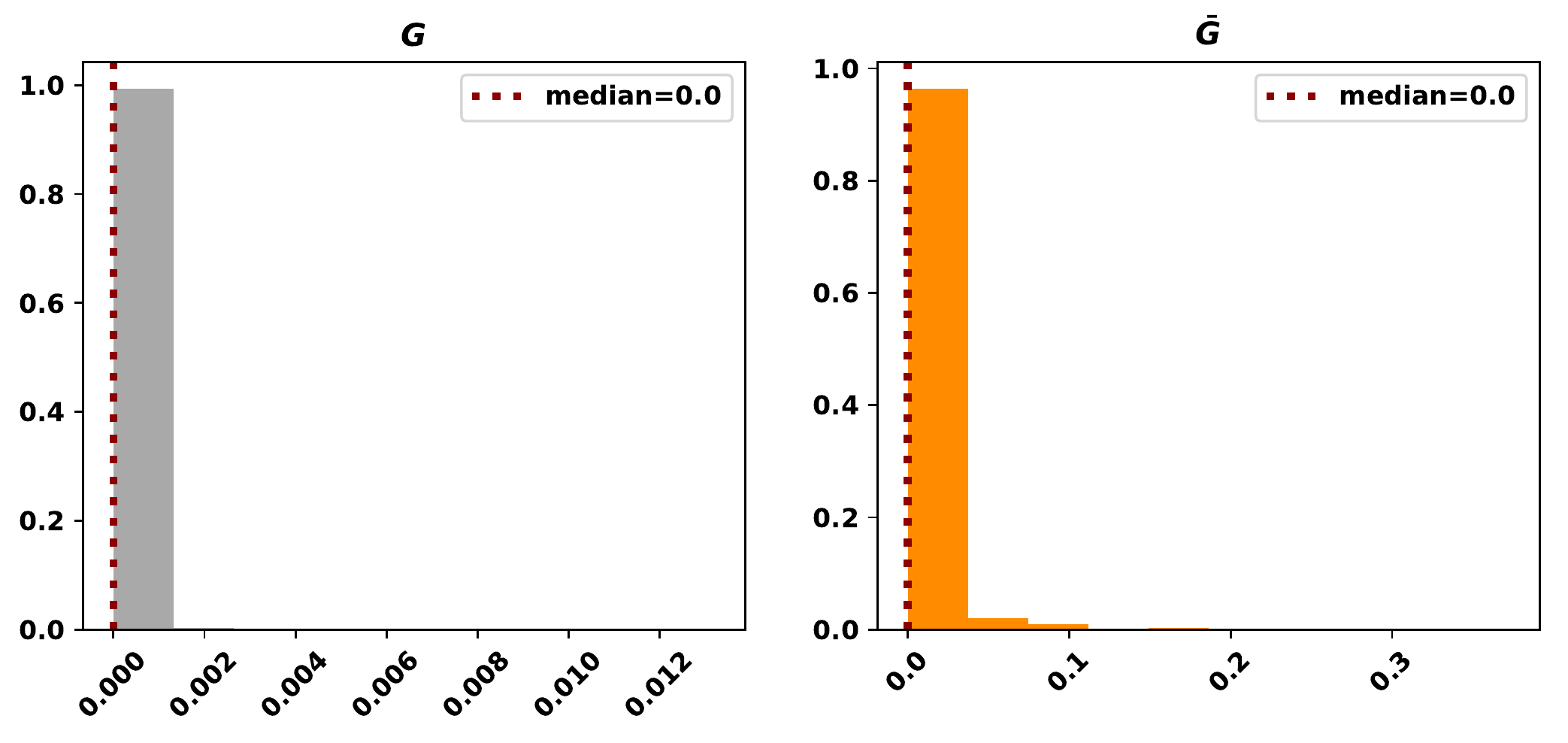}
    \caption{Distributions of betweenness centrality in \(G\) and \(\bar{G}\)}
    \label{fig:bc_distributions}
\end{figure}

\begin{figure}[!hbtp]
    \centering
    \includegraphics[width=.8\textwidth]{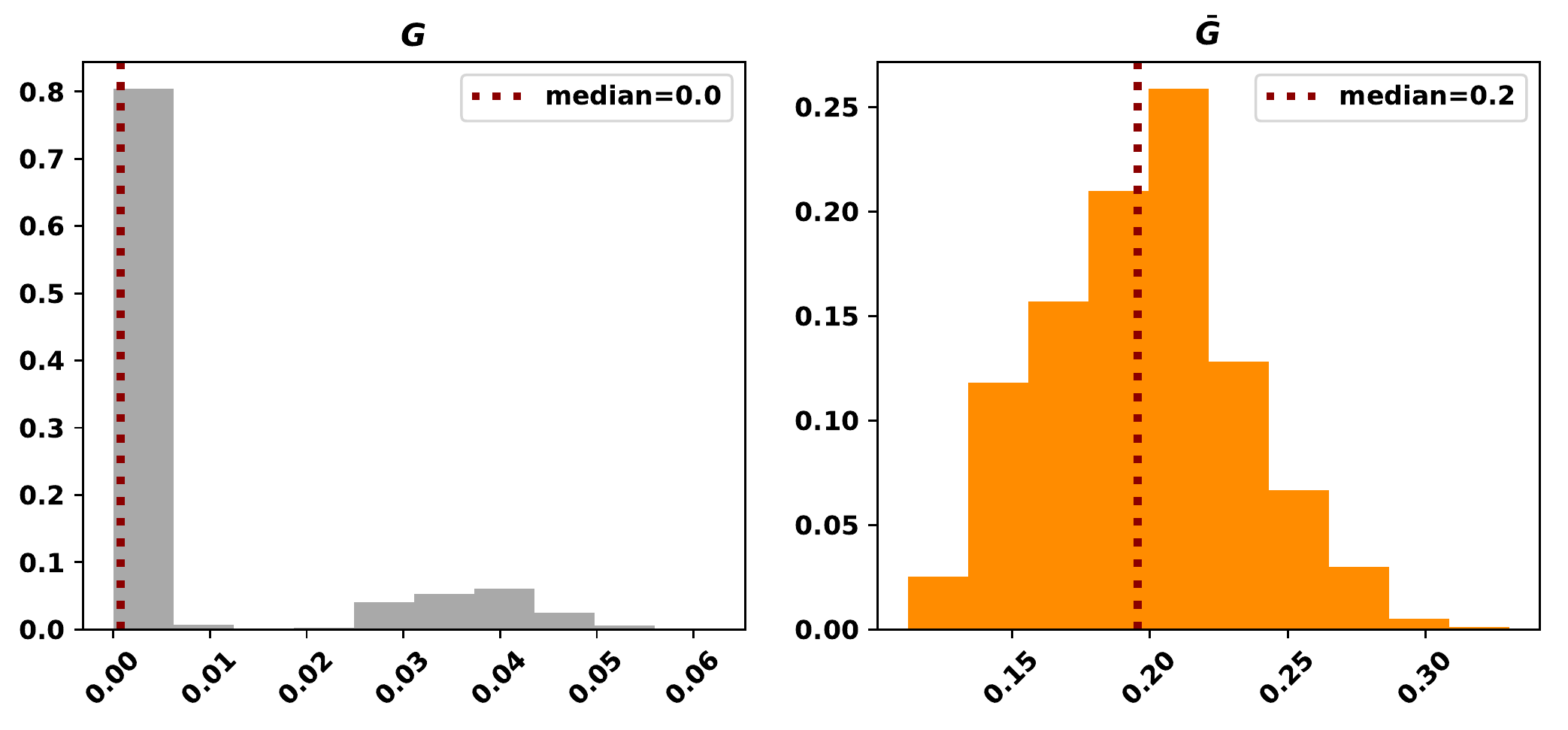}
    \caption{Distributions of closeness centrality in \(G\) and \(\bar{G}\)}
    \label{fig:cc_distributions}
\end{figure}

\subsection{Distrubutions for Topic Networks}\label{appendix:distributions}

\begin{figure}[!hbtp]
    \centering
    \includegraphics[width=\textwidth]{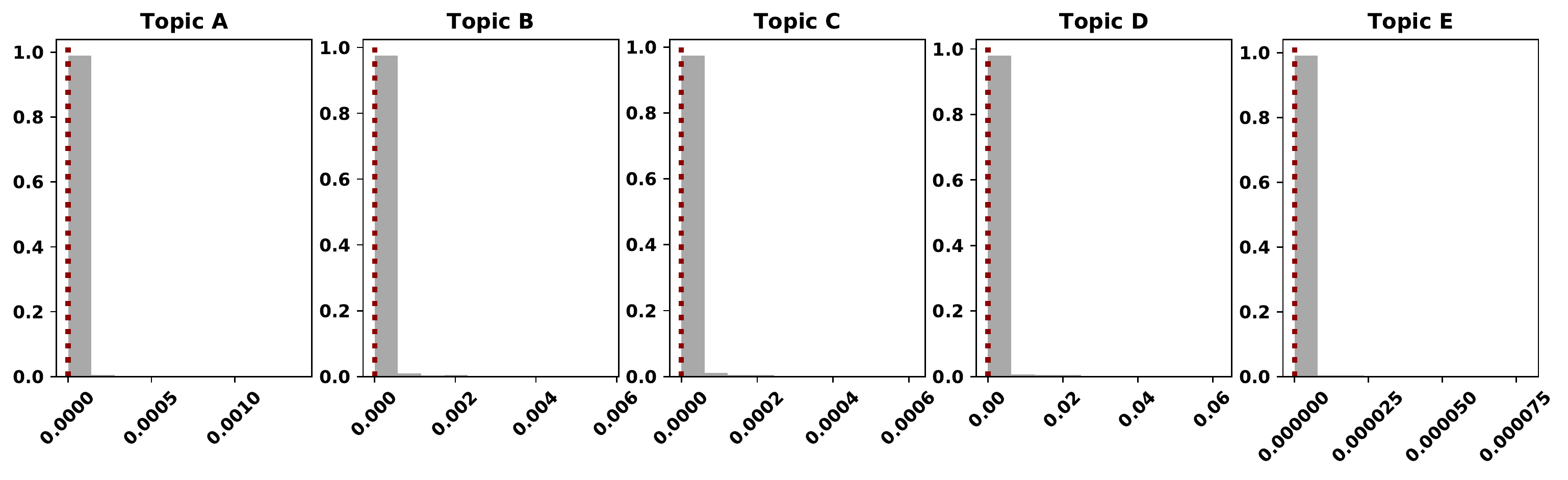}
    \caption{Distributions of betweenness centrality in topics' networks.}
    \label{fig:bc_distributions_topics}
\end{figure}

\begin{figure}[!hbtp]
    \centering
    \includegraphics[width=\textwidth]{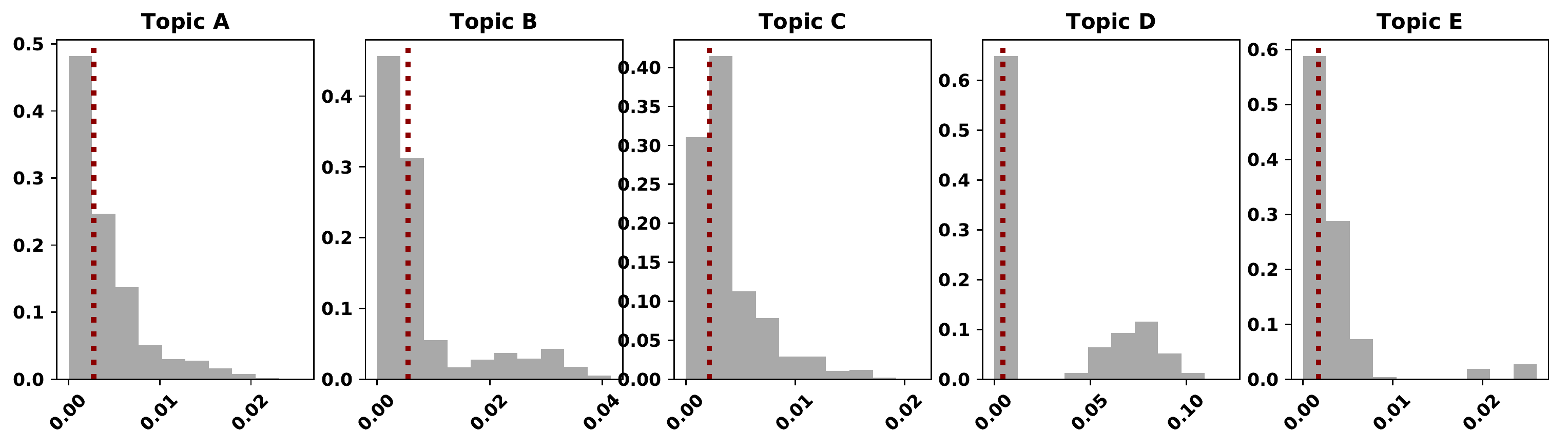}
    \caption{Distributions of closeness centrality in topics' networks.}
    \label{fig:cc_distributions_topics}
\end{figure}

\end{document}